\documentclass[a4,usenatbib,fleqn]{mn2e}
\usepackage{mathptmx}
\usepackage{amsmath,amssymb,amsbsy}
\usepackage[pdftex]{graphicx}
\usepackage[backref=none]{hyperref}
\usepackage[usenames,dvipsnames]{color}
\usepackage{bm}
\usepackage{txfonts}
\usepackage{graphicx}
\usepackage{lscape}   
\usepackage{threeparttable} 
\usepackage{dcolumn}

\DeclareSymbolFont{cmletters}{OML}{cmm}{m}{it}
\DeclareMathSymbol{v}{\mathalpha}{cmletters}{"76}
\def\gcc{\hbox{\rm\hskip.35em  g cm}$^{-3}$}
\def\rads{\hbox{\rm\hskip.35em rad s}$^{-1}$}
\def\radss{\hbox{\rm\hskip.35em  rad s}$^{-2}$}
\def\cms{\hbox{\rm\hskip.35em  cm s}$^{-1}$}

\def\ergs{\hbox{\rm\hskip.35em  erg s}$^{-1}$}

\def\yr{\hbox{\rm\hskip.35em yr}$^{-1}$}
\def\mevs{\hbox{\rm\hskip.35em  MeV$^{-1}$ s}$^{-1}$}
\def\hzs{\hbox{\rm\hskip.35em  Hz s}$^{-1}$}
\def\hzss{\hbox{\rm\hskip.35em  Hz s}$^{-2}$}

\definecolor{MyDarkBlue}{rgb}{0,0.1,0.7}
\hypersetup{pdfborder={0 0 0},colorlinks,breaklinks=true,
  urlcolor={blue},citecolor={MyDarkBlue},linkcolor={magenta}}

\newcommand{\abs}[1]{{\left|#1\right|}}

\newcommand{\apj}{ApJ}
\newcommand{\apjl}{ApJ}
\newcommand{\apjs}{ApJS}

\newcommand{\mnras}{MNRAS}
\newcommand{\nat}{Nature}

\newcommand{\aap}{A{\&}A}
\newcommand{\araa}{ARA{\&}A}

\newcommand{\apss}{Ap{\&}SS}
\newcommand{\arns}{Ann. Rev. Nucl. Sci}

\newcommand{\pasa}{PASA}
\newcommand{\pasj}{PASJ}

\newcommand{\prc}{Phys. Rev. C}
\newcommand{\prd}{Phys. Rev. D}
\newcommand{\prl}{Phys. Rev. Lett.}

\newcommand{\raa}{Res. Astron. Astrophys.}

\title[Glitches in gamma-ray pulsars]{Glitches in four gamma-ray pulsars and inferences on the neutron star structure}

\author[G\"{u}gercino\u{g}lu et al.]{E. G\"{u}gercino\u{g}lu$^{1,2}$\thanks{Contact e-mail: \href{mailto:egugercinoglu@gmail.com}{egugercinoglu@gmail.com}}, M.~Y. Ge$^{3}$\thanks{Contact e-mail: \href{mailto:gemy@ihep.ac.cn}{gemy@ihep.ac.cn}}, J.~P. Yuan$^{4,5}$, S.~Q. Zhou$^{6}$
\\
$^{1}$Hekim Tahsin Street No:7, Sar{\i}yer, 34467 Istanbul, Turkey \\
$^{2}$Faculty of Engineering and Natural Sciences, Sabanc{\i} University, Orhanl{\i}, Tuzla, 34956 Istanbul, Turkey \thanks{Former institution.} \\
$^{3}$Key Laboratory of Particle Astrophysics, Institute of High Energy Physics, Chinese Academy of Sciences, Beijing 100049, China \\
$^{4}$Xinjiang Astronomical Observatory, Chinese Academy of Sciences, Xinjiang 830011, China \\
$^{5}$Center for Astronomical Mega-Science, Chinese Academy of Sciences, Beijing, 100012, China \\
$^{6}$School of Physics and Astronomy, Sun Yat-Sen University, Zhuhai, 519082, China}

\date{Accepted 2022 January 3. Received 2021 December 16; in original form 2020 November 30}

\pubyear{2022}

\begin{document}
\label{firstpage}
\pagerange{\pageref{firstpage}--\pageref{lastpage}}
\maketitle

\begin{abstract}
We present timing solutions from the Fermi-LAT observations of gamma-ray pulsars PSR J0835$-$4510 (Vela), PSR J1023$-$5746, PSR J2111$+$4606, and PSR J2229$+$6114. Data ranges for each pulsar extend over a decade. From data analysis we have identified a total of 20 glitches, 11 of which are new discoveries. Among them, 15 glitches are large ones with $\Delta\nu/\nu\gtrsim10^{-6}$. PSR J1023$-$5746 is the most active pulsar with glitch activity parameter being $A_{\rm g}=14.5\times10^{-7}$\yr~in the considered data span and should be a target for frequently glitching Vela-like pulsars in future observations. We have done fits within the framework of the vortex creep model for 16 glitches with $\Delta\nu/\nu\gtrsim10^{-7}$. By theoretical analysis of these glitches we are able to obtain important information on the structure of neutron star, including moments of inertia of the superfluid regions participated in glitches and coupling time-scales between various stellar components. The theoretical prediction for the time to the next glitch from the parameters of the previous one is found to be in qualitative agreement with the observed inter-glitch time-scales for the considered sample. Recoupling time-scales of the crustal superfluid are within the range of theoretical expectations and scale inversely with the spin-down rate of a pulsar. We also determined a braking index n=2.63(30) for PSR J2229$+$6114 after glitch induced contributions have been removed.  
\end{abstract}





\section{Introduction} \label{sec:intro}

Pulsars are neutron stars powered by their fast rotation and huge magnetic fields, and are capable of producing observed pulsed radiation from radio to $\gamma$ rays by accelerating charged particles in the magnetosphere to ultra-relativistic speeds with large Lorentz factors \citep{ruderman72,srinivasan89,spitkovsky06,beskin18}. The origin of their short duration, stable pulses is still one of the mysteries of pulsar astrophysics and is explained in terms of plasma effects forming rotating lighthouse-like beams sweeping the observer at the angular velocity, $\Omega=2\pi\nu$, of the underlying neutron star with $\nu$ being the rotational frequency. Such emission can be produced at low altitudes just above the pulsar polar cap \citep{ruderman75,daugherty96} or further out in the magnetosphere near the light cylinder \citep{arons83,cheng86,romani96,dyks03} at so-called gaps.  

Information on the neutron star internal structure and dynamics is obtained through pulsar timing and spectral observations \citep{baym75, haskell18}. Certain types of timing variability in pulsars are puzzling and subject of intense study \citep{dalessandro96,parthasarathy19}. Pulsars slow down steadily via magneto-dipole radiation so that their rotational evolution is predictable once their spin frequency $\nu$, spin-down rate $\dot\nu$ and frequency second time derivative $\ddot\nu$ are measured precisely. Nevertheless, such an evolution is occasionally interrupted by glitches \citep{yuan10,espinoza11,shabanova13,yu13,basu20,basu21,liu21,lower21}. Glitches manifest themselves as sudden increase in the rotation rate of pulsars with relative sizes $\Delta\nu/\nu\sim10^{-10}-10^{-5}$, and are usually accompanied by increments in the magnitude of the spin-down-rate with still greater ratios $\Delta\dot\nu/\dot\nu\sim10^{-4}-10^{-1}$. To date over six hundred glitches have been detected in about two hundred pulsars\footnote{See continually updated Jodrell Bank glitch catalogue at the URL: www.jb.man.ac.uk/pulsar/glitches/gTable.html \citep{espinoza11}}. Post-glitch recovery includes initial exponential decay(s) followed by long-term relaxation of the increase in the spin-down rate and possible permanent increases in both frequency and spin-down rate that do not heal at all. Relaxation towards the original pre-glitch conditions occurs on time-scales ranging from several days to a few years. The implied rapid angular momentum transfer and very long recovery demonstrate that neutron stars contain interior superfluid component(s) \citep{baym69}. Observations of glitches, thus, provide valuable information on the characteristics of the superfluid regions participating in glitches and their coupling with various stellar components \citep{sauls89,haskell15}. Neutron star crust \citep{alpar89,link96,erbil20} and core region \citep{ruderman98,erbil17,sourie20} may respond to glitch induced changes and take part in the long-term recovery process.

Given their young ages and high spin-down rates, gamma-ray pulsars are good targets for studying glitches \citep{sokolova20}. Radio-quiet gamma-ray pulsar PSR J2021+4026 has displayed repeated transitions from high $\gamma$-ray /low spin-down rate state to low $\gamma$-ray /high spin-down rate level \citep{takata20}. This $\gamma$-ray flux variability coinciding with persistent spin-down rate changes may result from glitches. Therefore, observations of glitches in gamma-ray pulsars have potential to provide rich information on both magnetospheric processes and internal structure, bridging the gap between neutron star exterior and interior. In this paper, we present timing solutions of 20 glitches observed in four gamma ray pulsars by the Fermi-LAT and examine their properties in terms of the vortex creep model. 

The paper is organised as follows. In \S\ref{sec:observations} we summarize observations and data analysis. In \S\ref{sec:creepmodel}, main observational properties of glitches are presented and the vortex creep model is outlined. In \S\ref{sec:glitches}, we report on the glitches identified from data analysis and give fits for large glitches done with the vortex creep model. In \S\ref{sec:results}, information on the neutron star structure extracted from post-glitch timing fits is given. We discuss our findings in \S\ref{sec:dandc}. While \S\ref{sec:comparison} includes the properties of other glitching gamma-ray pulsars published in the literature.

\section{Observations and Data Analysis} \label{sec:observations}

Studies of $\gamma$-ray emission from pulsars have increased significantly after
the launch of the Fermi satellite carrying the Large Area Telescope
(LAT) in 2008, which has a sensitivity from 20 MeV to around 300
GeV (for a review, see \cite{caraveo14}). A
multitude of gamma-ray pulsars has been discovered
thereafter thanks to its wide field of view covering 2.4
sr which led to continuously mapping of the whole sky \citep{abdo10,abdollahi20}
This all sky survey is completed in about 3
h. So, observations of several weeks to a few months will
suffice to build an average pulse profile for confirming an object as a
gamma-ray source. Photons are collected to an accuracy better than
1$\mu$s, so that even rapid pulsations from millisecond pulsars can be
easily detected. This precise and regular monitoring resulted in the discovery of several radio-quiet
pulsars and many follow-up observations of radio-loud gamma ray
pulsars, amounting to hundreds of sources in total.

In this work, we use the LAT data (from 2008 August 4 to 2021 September 16) to perform timing analysis,
which are analysed with the Fermi Science Tools (v10r0p5)
\footnote{https://fermi.gsfc.nasa.gov/ssc/data/analysis/scitools/pulsar\_analysis\_tutorial.html}.
The events are selected with the angular distance less than 3\,\textordmasculine
\,of the Vela pulsar and a zenith angle of less than 105\,\textordmasculine\,and
energy range 0.03 to 10\,GeV \citep{abdo09,abdo10b}. After event selection, the
arrival time of each event is corrected to SSB with DE405 using {\it gtbary}. To obtain Time of Arrivals (ToAs), $\nu$ and $\dot{\nu}$ of each pulsar are determined by folding the counts to 
reach the maximum Pearson $\chi^{2}$ \citep{ge12}. After this step, each ToA is
accumulated from several days exposure. According to the flux, the accumulated time for PSRs J0835--4510 (Vela),
J1023--5746, J2111+4606, and J2229+6114 are 1, 10, 8 and 30-day exposure, respectively.
Then, the spin evolution and glitch parameters are fitted by 
\begin{eqnarray}
\nu=&\nu_{0}+\dot\nu_{0}{(t-t_{0})}+\frac{1}{2}\ddot\nu_{0}{(t-t_{0})^{2} + \Delta\nu_{\rm g}+\Delta\dot\nu_{\rm g}}(t-t_{\rm g})\nonumber\\&+\Delta\nu_{\rm d}\exp(-(t-t_{\rm g})/\tau_{\rm d}),
\label{eq11}
\end{eqnarray}
from the ToAs via TEMPO2 \citep{hobbs06}. Here $\nu_{0}$, $\dot\nu_{0}$ and $\ddot\nu_{0}$ are spin parameters measured at epoch $t_{0}$. 
$\Delta\nu_{\rm g}$ and $\Delta\dot\nu_{\rm g}$ are the jump values of the spin frequency and its derivative at $t_{\rm g}$.
$\Delta\nu_{\rm d}$ and $\tau_{d}$ are the amplitude and decay time of exponential decay component of glitch.

\begin{figure}
\centering
\vspace{0.1cm}
\includegraphics[width=1.0\linewidth]{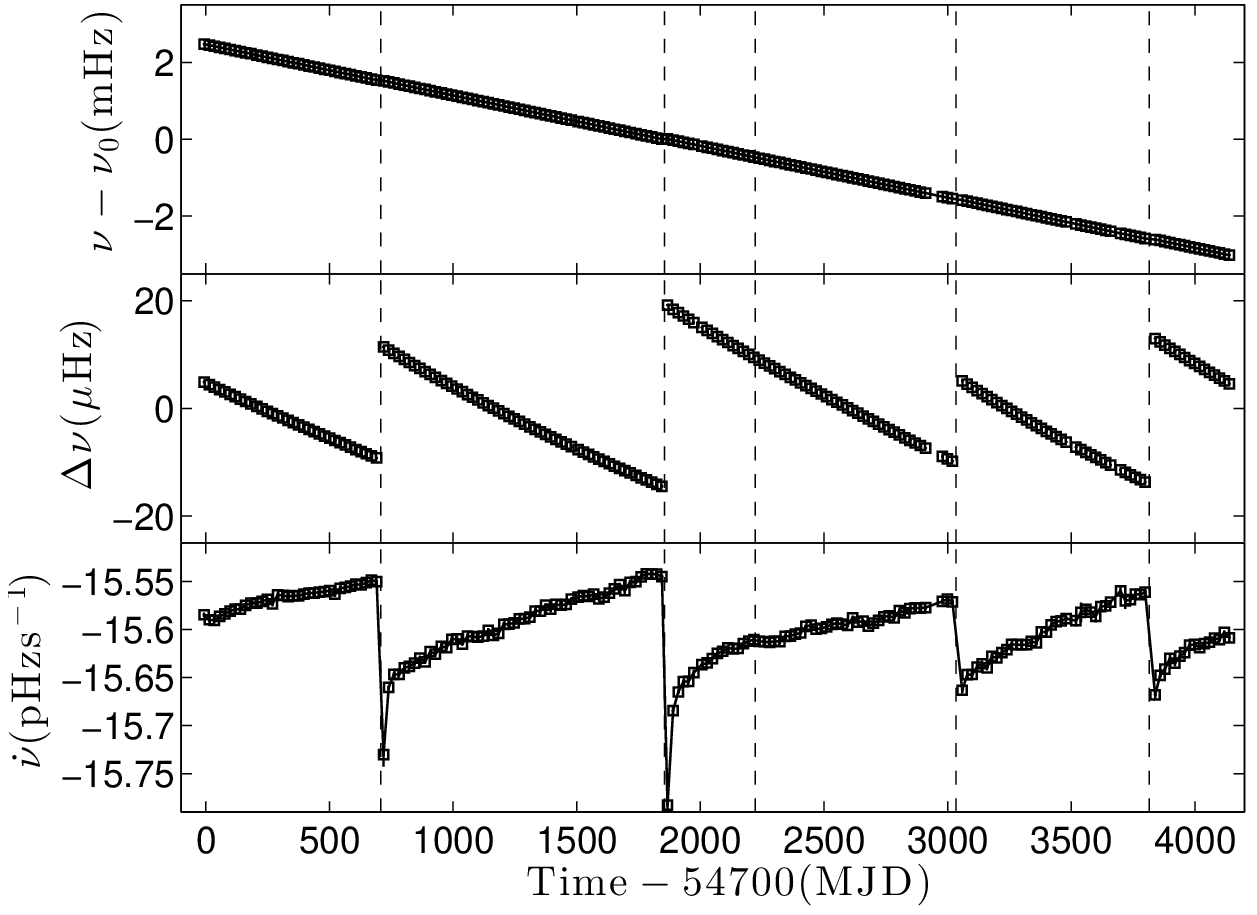}
\caption{The spin evolution of the Vela pulsar (PSR J0835$-$4510) in the range MJD 54692-58839. The top panel shows the frequency evolution with respect to fidicual value $\nu_{0}=11.19$ Hz. The middle panel shows the pulse-frequency residuals $\Delta\nu$, obtained by subtraction of the pre-glitch timing solution. The bottom panel shows the variation of the spin-down rate. The glitch epoches are indicated by vertical dashed lines.}
\label{J0835spin}
\end{figure}

PSR J0835--4510 was discovered in the Vela supernova remnant in 1967 \citep{large68}. The Vela pulsar is a radio-loud pulsar with spin-down rate $\abs{\dot\nu}=1.56\times10^{-11}$\hzs, spin-down age $t_{\rm sd}=11.3$ kyr, spin-down power $\dot E_{\rm sd}=6.9\times10^{36}$\ergs, surface magnetic field $B_{\rm s}=3.4\times10^{12}$ G, and magnetic field strength at light cylinder $B_{\rm LC}=44.5$ kG. The long term spin evolution of the Vela for the interval MJD 54692-58839 analysed in this paper is shown in Fig. \ref{J0835spin}. 

\begin{figure}
\centering
\vspace{0.1cm}
\includegraphics[width=1.0\linewidth]{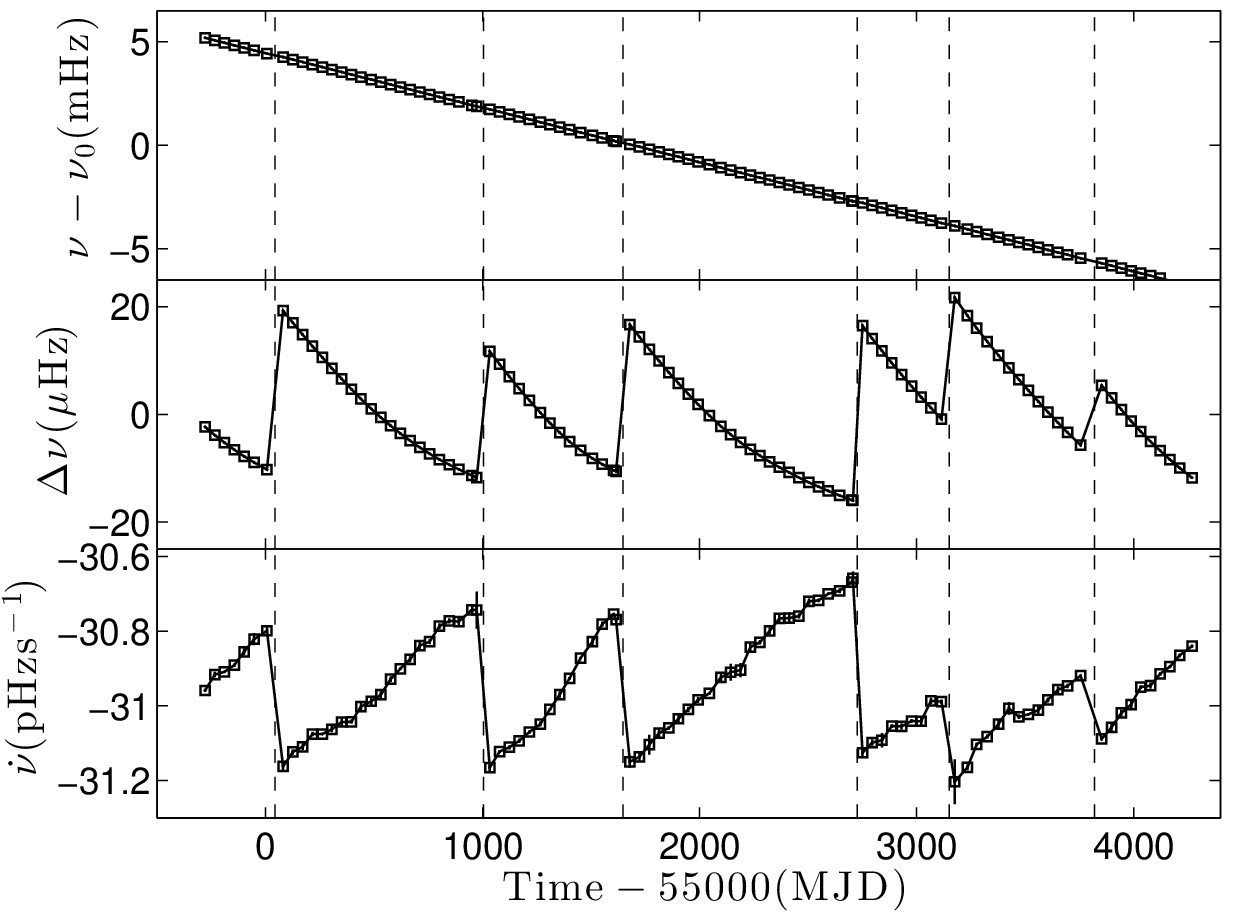}
\caption{The spin evolution of PSR J1023$-$5746 in the range MJD 54728-59269. The top panel shows the frequency evolution with respect to fidicual value $\nu_{0}=8.971$ Hz. The middle panel shows the pulse-frequency residuals $\Delta\nu$, obtained by subtraction of the pre-glitch timing solution. The bottom panel shows the variation of the spin-down rate. The glitch epoches are indicated by vertical dashed lines.}
\label{J1023spin}
\end{figure}

PSR J1023--5746 was discovered by the Fermi-LAT in 2009 \citep{sazparkinson10}. PSR J1023$-$5746 is a radio-quiet pulsar with spin-down rate $\abs{\dot\nu}=3.09\times10^{-11}$\hzs, spin-down age $t_{\rm sd}=4.6$ kyr, spin-down power $\dot E_{\rm sd}=1.1\times10^{37}$\ergs, surface magnetic field $B_{\rm s}=6.6\times10^{12}$ G, and magnetic field strength at light cylinder $B_{\rm LC}=44.8$ kG. The long term spin evolution of PSR J1023--5746  for the interval MJD 54728-59269 analysed in this paper is shown in Fig. \ref{J1023spin}. 

\begin{figure}
\centering
\vspace{0.1cm}
\includegraphics[width=1.0\linewidth]{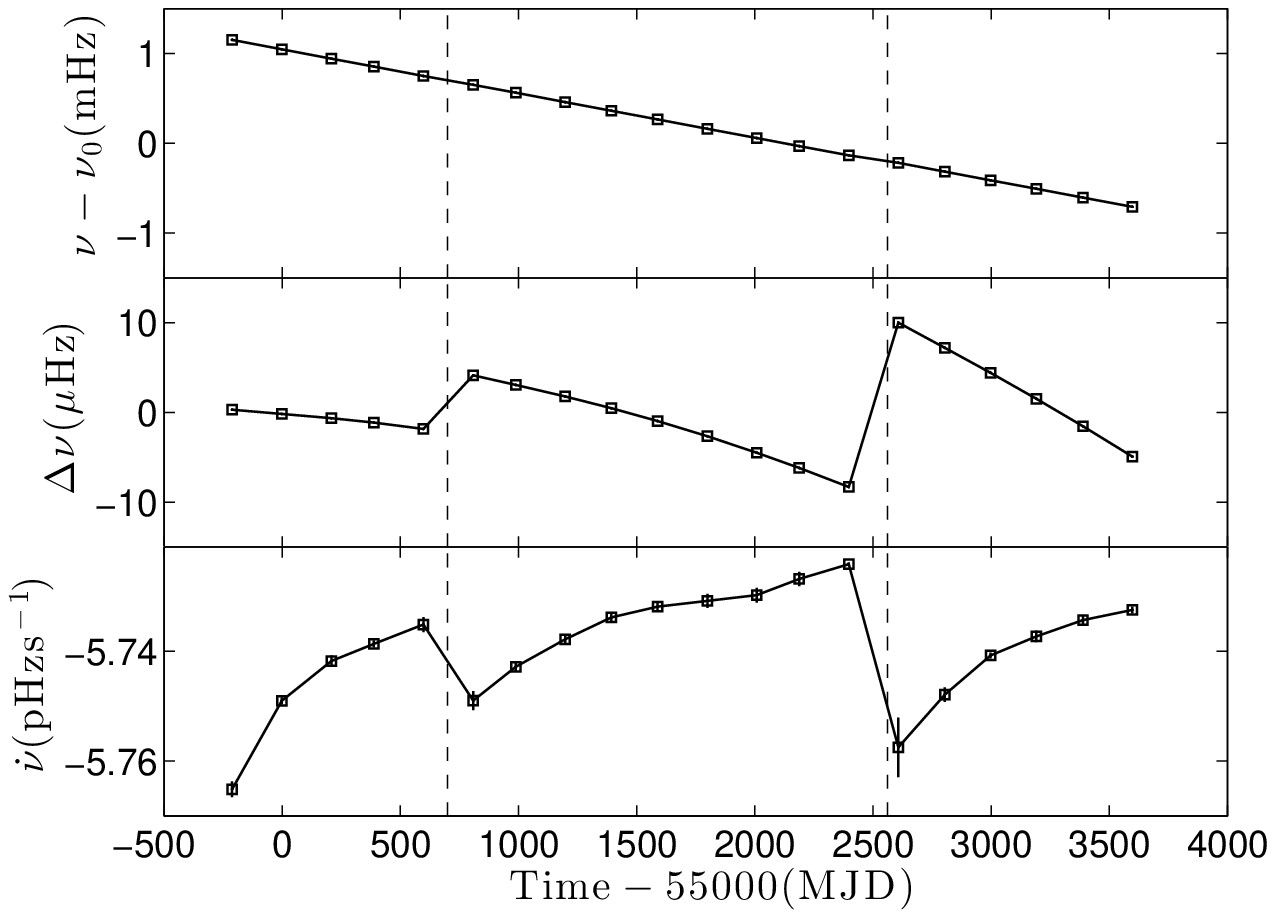}
\caption{The spin evolution of PSR J2111$+$4606 in the range MJD 54787-58598. The top panel shows the frequency evolution with respect to fidicual value $\nu_{0}=6.336$ Hz. The middle panel shows the pulse-frequency residuals $\Delta\nu$, obtained by subtraction of the pre-glitch timing solution. The bottom panel shows the variation of the spin-down rate. The glitch epoches are indicated by vertical dashed lines.}
\label{J2111spin}
\end{figure}

PSR J2111$+$4606 was discovered by the Fermi-LAT in 2009 \citep{pletsch12}. PSR J2111$+$4606 is a radio-quiet pulsar with spin-down rate $\abs{\dot\nu}=5.7\times10^{-12}$\hzs, spin-down age $t_{\rm sd}=17.5$ kyr, spin-down power $\dot E_{\rm sd}=1.4\times10^{36}$\ergs, surface magnetic field $B_{\rm s}=4.8\times10^{12}$ G, and magnetic field strength at light cylinder $B_{\rm LC}=11.5$ kG. The long term spin evolution of PSR J2111$+$4606  for the interval MJD 54787-58598 analysed in this paper is shown in Fig. \ref{J2111spin}.

\begin{figure}
\centering
\vspace{0.1cm}
\includegraphics[width=1.0\linewidth]{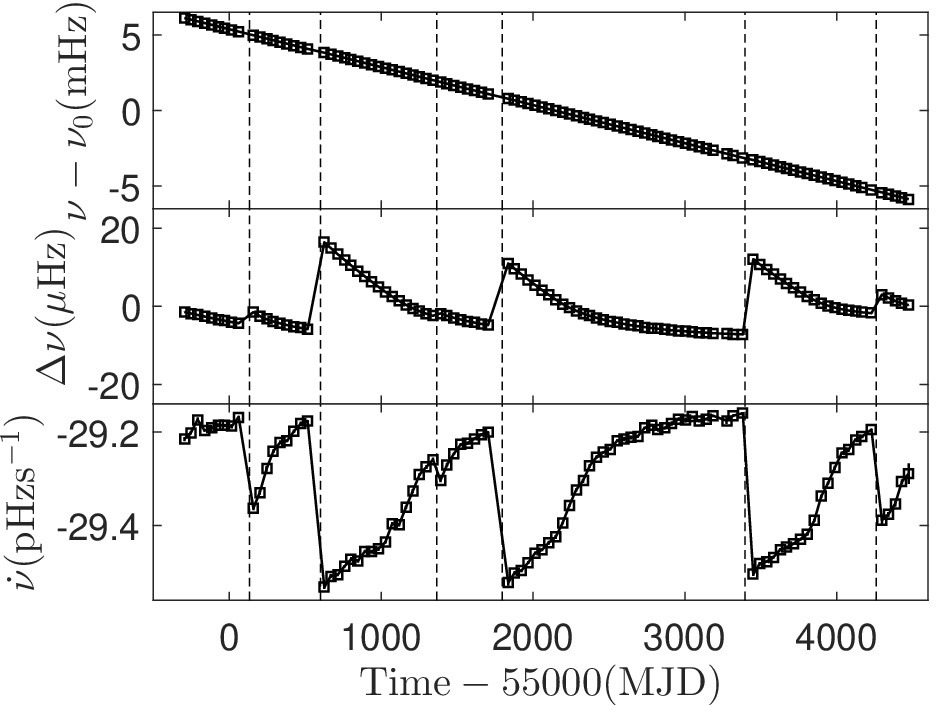}
\caption{The spin evolution of PSR J2229$+$6114 in the range MJD 54705-59473. The top panel shows the frequency evolution with respect to fidicual value $\nu_{0}=19.36$ Hz. The middle panel shows the pulse-frequency residuals $\Delta\nu$, obtained by subtraction of the pre-glitch timing solution. The bottom panel shows the variation of the spin-down rate. The glitch epoches are indicated by vertical dashed lines.}
\label{J2229spin}
\end{figure}

PSR J2229$+$6114 was discovered by \citet{halpern01} in 2001. PSR J2229$+$6114 is a radio-loud pulsar with spin-down rate $\abs{\dot\nu}=2.94\times10^{-11}$\hzs spin-down age $t_{\rm sd}=10.5$ kyr, spin-down power $\dot E_{\rm sd}=2.2\times10^{37}$\ergs, surface magnetic field $B_{\rm s}=2.03\times10^{12}$ G, and magnetic field strength at light cylinder $B_{\rm LC}=139$ kG. The long term spin evolution of PSR J2229$+$6114  for the interval MJD 54705-59473 analysed in this paper is shown in Fig. \ref{J2229spin}.

Observational cadence for pulsars in our sample is not same for each source. For example PSR J1023$-$5746 is observed once in about 45 days with 98 data points in total. While for PSR J2111$-$4606 observations are made in every four months, resulting in only 20 data points.
\section{Glitch Characteristics and Post-glitch Recovery in Terms of the Vortex Creep Model } \label{sec:creepmodel}

Glitches are observed from pulsars of all ages but they are predominantly seen from young and mature pulsars with ages $10^{3}-10^{5}$ yr \citep{espinoza11,yu13}. Glitch sizes $\Delta\nu $, which extends from $10^{-4}\mu$Hz to about $100 \mu$Hz, show a bimodal distribution with a sharp peak at $18 \mu$Hz for large glitches and a broader dispersion for small glitches becoming maximal around $\sim3\times10^{-2}\mu$Hz \citep{ashton17,fuentes17,basu21}. 

A pulsar's glitching rate is quantified by glitch activity parameter $A_{\rm g}$. It is defined as the total change in the frequency of a pulsar due to glitches within a total observation time $T_{\rm obs}$ and is given by \citep{mckenna90}
\begin{equation}
A_{\rm g}=\frac{1}{T_{\rm obs}}\sum_{i=1}^{N_{\rm g}} \left(\frac{\Delta\nu}{\nu}\right)_{i}.
\label{gactivity}
\end{equation}
If glitches involve whole crustal superfluid and angular momentum reservoir is depleted at regular intervals for a given pulsar, then Eq. (\ref{gactivity}) constrains the equation of state (EOS) of neutron star matter \citep{datta93,link99} and in turn can be used to determine the pulsar mass \citep{ho15,pizzochero17}.

A superfluid system achieves rotation by forming quantized vortex lines. In the neutron star crust, vortex lines coexist with lattice nuclei. These nuclei provide attractive pinning sites to vortex lines. Note, however, that at shallow densities vortex-nucleus interaction is repulsive and interstitial pinning occurs \citep{alpar77,link91,antonelli20}. The distribution of vortices is nonuniform due to the presence of crustal inhomogeneities like cracks, dislocations and impurities formed after quakes and results in a configuration in which high vortex density regions are surrounded by vortex depletion regions, collectively dubbed as vortex traps \citep{cheng88}. Glitches are thought to be arising from collective vortex unpinning events induced in traps when angular velocity lag $\omega\equiv\Omega_{\rm s}-\Omega_{\rm c}$ between the superfluid and the crustal normal matter exceeds a critical threshold \citep{anderson75,cheng88}. Following unpinning, vortex lines impart their angular momentum to the crust via coupling of kelvon wave excitations to the lattice phonons {\bf \citep{epstein92,jones92,graber18}}. This mechanism leads to quite short glitch rise times in agreement with the current observational upper limit 12.6 seconds obtained for the spin-up time-scale of the 2016 Vela glitch \citep{ashton19}. In between glitches, finite temperature of the neutron star crust impedes perfect pinning which causes vortex lines to overcome pinning barriers sustained by nuclei and skip many pinning sites. In a slowing down neutron star, this thermally activated vortex motion, ``creep'' \citep{alpar84}, bring about radially outward migration of vortices and thereby allows superfluid to spin down in order to keep up with rotation rate of the crust. In the aftermath of a glitch, the steady state lag abruptly declines which leads to a complete stoppage of the creep in some superfluid regions. Slow post-glitch recovery reflects the healing of the creep until steady-state conditions are re-established by external braking torque.  

According to the vortex creep model, formation, short time-scale relaxation, and long term recovery phases of glitches proceed as follows \citep{alpar84,erbil20}:
\begin{enumerate}
\item A crustquake leads to the vortex trap formation and triggers vortex unpinning avalanche. Unpinned vortices transfer their excess angular momentum to the crust, and spin it up which is observed as a glitch. 
\item Crust breaking may leave behind ephemeral magnetospheric signatures coinciding with the glitch date like pulse shape, polarization, and emission changes which are associated with plastic motion of the magnetic field lines anchored to the broken platelet. Such changes are resolved for the glitches of the Vela \citep{palfreyman18}, the Crab \citep{feng20}, PSR J1119$-$6127 \citep{weltevrede11}, and other pulsars \citep{yuan19}. 
\item Parts of the superfluid wherein unpinning has taken place decouple from the external decelerating torque. Since external torque is now acting on less moment of inertia, spin-down rate undergoes a step increase. 
\item If vortex unpinning event includes a vortex inward motion due to platelet movement, the glitch magnitude is expected to attain its maximum value after some time has elapsed \citep{erbil19}. Such delayed spin-ups are observed after the Crab glitches \citep{wong01,shaw18,ge20}. 
\item Immeadiately after the large-scale vortex unpinning avalanche, the excess angular momentum of the crustal superfluid is first shared with the normal matter nuclei and electrons in the crust which leads to a peak in the crustal rotation rate following a glitch. The change in the  crustal rotational state is then communicated to the core superfluid via electrons' scattering off magnetised vortex lines \citep{ALS84}. Therefore, initial peak in the observed crustal rotation rate relaxes on the crust-core coupling time-scale. After that post-glitch equilibrium rotation rate is attained. A peak in the rotation rate of the normal component will occur if the coupling of the crustal superfluid with normal matter is stronger than its coupling with the core superfluid \citep{pizzochero20}. This is the observed case after the Vela glitches \citep{dodson02,dodson07,ashton19}.
\item After the core superfluid come into equilibrium with the normal matter in the crust, they behave like a single component which comprises most of the moment of inertia of the neutron star. Then, the only component decoupled from the braking torque on longer time-scales is the crustal superfluid and slow post-glitch evolution reflects the response of crustal superfluid to the glitch induced changes.
\end{enumerate}

In the vortex creep model, post-glitch spin-down rate behaviour mimics responses of the vortex creep regions either in linear or non-linear regimes to a glitch. In certain parts of the superfluid in the inner crust and the outer core vortex creep is linear in glitch induced changes, leading to exponential relaxation \citep{alpar89,erbil17}:
\begin{align}
\Delta\dot\nu(t)=&-\frac{I_{\rm exp}}{I_{\rm c}}\frac{\Delta\nu}{\tau_{\rm exp}}e^{-t/\tau_{\rm exp}}.
\label{creepexp}
\end{align}

The exponential decay time-scale associated with a linear creep region in the crustal superfluid is \citep{alpar89}
\begin{equation}
\tau_{\rm lin} = \frac{kT }{E_{\rm p}} \frac{R \omega_{\rm cr}}{4 \Omega_{\rm s} v_{0}} \exp \left( \frac{E_{\rm p}}{kT} \right),
\label{taulin}
\end{equation}
where $k$ is the Boltzmann constant, $T$ is the temperature, $E_{\rm p}$ is the pinning energy between vortex line and nucleus, $\Omega_{\rm s}$ is the superfluid angular velocity, $\omega_{\rm cr}$ is the critical lag for unpinning, and $R$ is the distance of vortex lines from the rotation axis. Microscopic vortex velocity around nuclei amounts to $v_{0}=10^{5}-10^{7}$\cms~throughout the crust \citep{erbil16}.

In the outer core, vortex creep against flux tubes also gives rise to exponential type relaxation with a decay time-scale \citep{erbil14}
\begin{align}
\tau_{\rm tor}\simeq60&\left(\frac{\vert\dot{\Omega}\vert}{10^{-10}\mbox{
\radss}}\right)^{-1}\left(\frac{T}{10^{8}\mbox{
K}}\right)\left(\frac{R}{10^{6}\mbox{ cm}}\right)^{-1}
x_{\rm p}^{1/2}\times\nonumber\\&\left(\frac{m_{\rm p}^*}{m_{\rm p}}\right)^{-1/2}\left(\frac{\rho}{10^{14}\mbox{\gcc}}\right)^{-1/2}\left(\frac{B_{\phi}}{10^{14}\mbox{ G}}\right)^{1/2}\mbox{days,}
\label{tautor}
\end{align}
where $\dot{\Omega}$ is the spin-down rate, $x_{\rm p}$ is the proton fraction in the neutron star core, $\rho$ is the matter density, $ m_{\rm p}^*(m_{\rm p})$ is the effective (bare) mass of  protons, and $B_{\phi}$ is the toroidal magnetic field strength.

For a single crustal superfluid region in the non-linear regime, the post-glitch response is \citep{alpar84}
\begin{equation}
\Delta\dot\nu (t)=\frac{I_{\rm s}}{I_{\rm c}}\dot\nu_{0}\left[1-\frac{1}{1+({\rm e}^{t_{\rm s}/\tau_{\rm s}}-1){\rm e}^{-t/\tau_{\rm s}}}\right],
\label{creepsingle}
\end{equation}
and has the form of making a shoulder such that at about $t_{\rm s}$ after the glitch the spin-down rate decreases rapidly by $I_{\rm s}/I_{\rm c}$ in a time $\sim3\tau_{\rm s}$ quasi exponentially. This response resembles the behaviour of the Fermi distribution function of statistical mechanics for non-zero temperatures \citep{alpar84} and has been observed after several glitches \citep{buchner08,yu13,espinoza17}. Single superfluid region's recovery is thus completed on a time-scale
\begin{equation}
t_{\rm single}=t_{\rm s}+3\tau_{\rm s}.
\label{tgsingle}
\end{equation}

These non-linear creep regions are sites for the collective vortex unpinning avalanche. The time evolution of the crustal rotation rate after a glitch due to angular momentum transfer from non-linear creep regions takes the form \citep{erbil21}
\begin{equation}
\nu(t)=\nu(0)+\frac{I}{I_{\rm c}}\dot\nu_{0}t +\frac{I_{\rm s}}{I_{\rm c}}\varpi\ln \Bigg\{ 1+\frac{I}{I_{\rm c}}\exp\left(-\frac{t_{\rm s}}{\tau_{\rm s}}\right)\left[\exp\left(\frac{t}{\tau_{\rm s}}\right)-1\right]\Bigg\},
\end{equation}
\label{nunl}
where $\nu(0)=\nu_{0}+\Delta\nu$ is the rotation rate of the crust at the time of the glitch and $\varpi\equiv(kT/E_{\rm p})\omega_{\rm cr}$.

Eq.(\ref{creepsingle}) can be integrated explicitly by taking contributions of several neighbouring single non-linear regime regions into account with the assumption of linearly decreasing superfluid angular velocity during a glitch in a large extended crustal superfluid layer with radial extent $\delta r_{0}$, i.e. $\delta\Omega_{\rm s}(r, 0)=(1-r/\delta r_{0})\delta\Omega_{0}$ \citep{alpar96}

\begin{align}
\Delta\dot\nu(t)=\frac{I_{\rm A}}{I_{\rm c}}\dot\nu_{0}\left(1-\frac{1-(\tau_{\rm nl}/t_{0})\ln\left[1+(e^{t_{0}/\tau_{\rm nl}}-1)e^{-t/\tau_{\rm nl}}\right]}{1-e^{-t/\tau_{\rm nl}}}\right).
\label{creepfull}
\end{align}
Here total moment of inertia of the non-linear creep regions affected by vortex unpinning events is denoted by $I_{\rm A}$. In equations (\ref{creepsingle}) and (\ref{creepfull}) non-linear creep relaxation time is given by \citep{alpar84}
\begin{equation}
\tau_{\rm nl}=\tau_{\rm s}\equiv \frac{kT}{E_{\rm p}}\frac{\omega_{\rm cr}}{\vert\dot{\Omega}\vert}.
\label{taunl}
\end{equation}

For $\tau_{\rm nl}\lesssim t \lesssim t_{0}$ Eq.(\ref{creepfull}) reduces to
\begin{equation}
\frac{\Delta \dot \nu(t)}{\dot \nu}=\frac{I_{\rm A}}{I_{\rm c}}\left(1-\frac{t}{t_{0}}\right).
\label{glitchdotnu}
\end{equation}
Eq.(\ref{glitchdotnu}) leads to anomalously large inter-glitch $\ddot\nu$ and in turn very high braking index after the glitches
\begin{equation}
\ddot\nu=\frac{I_{\rm A}}{I_{\rm c}}\frac{\abs{\dot\nu_{0}}}{t_{0}},
\label{doubledotnu}
\end{equation}
which can be tested observationally \citep{yu13,dang20,lower21}.

In the integrated non-linear creep case described by Eq.(\ref{creepfull}), glitch induced changes in the spin-down rate recovers completely after a waiting time
\begin{equation}
t_{0}=\frac{\delta\omega}{\abs{\dot\Omega}}=\frac{\delta\Omega_{\rm s}+\Delta\Omega_{\rm c}}{\abs{\dot\Omega}}\cong\frac{\delta\Omega_{\rm s}}{\abs{\dot\Omega}},
\label{offsettime}
\end{equation}
with $\delta\Omega_{\rm s}$ being the decrement in the superfluid angular velocity due to vortex discharge and $\Delta\Omega_{\rm c}$ is the observed glitch magnitude. In the vortex creep model, post-glitch recovery is expected to end when the relaxations of the non-linear creep regions are over. So the theoretical expectation $t_{\rm th}=\mbox{max}(t_{0},t_{\rm single})$, i.e. maximum of Eq.(\ref{offsettime}) and Eq.(\ref{tgsingle}) gives a prediction for the inter-glitch time. 

Besides the non-linear creep regions giving rise to glitches via collective vortex unpinning with moments of inertia $I_{\rm s}$ and $I_{\rm A}$, there are vortex free regions surrounding the vortex traps with moment of inertia $I_{\rm B}$. As unpinned vortices traverse these regions at the time of the glitch, $I_{\rm B}$ can be determined from angular momentum balance
\begin{equation}
I_{\rm c}\Delta \Omega_{\rm c} = (I_{\rm A}/2 +I_{\rm s}+ I_{\rm B}) \delta \Omega_{\rm s},
\label{glitchomega}
\end{equation}
where the factor 1/2 accounts for linear decrease of $\delta \Omega_{\rm s}$ within $I_{\rm A}$.

In the neutron star crust dissipationless coupling of the free neutrons by lattice nuclei restricts their mobility and reduces the deposited superfluid angular momentum which is tapped at glitches \citep{chamel17}. Band theory calculations of \citet{chamel17} show that this entrainment effect may give rise to larger effective masses for neutrons with average enhancement factor $\langle m_{\rm n}^{*}/m_{\rm n}\rangle=5$ . Therefore, moments of inertia of the various superfluid components found from the post-glitch timing fits should be multiplied by the same enhancement factor. However, if the neutron star crustal lattice is disordered, then effects of the crustal entrainment are less pronounced \citep{sauls20}. Also some parts of the neutron star outer core may be involved in glitches \citep{erbil14,montoli20}, which lessens the restrictions brought by the crustal entrainment effect. 

The vortex creep model equations \citep{alpar84,alpar89} were derived in the Newtonian framework. While general relativistic (GR) effects can be important for abrupt angular momentum exchange process leading to short glitch rise times (e.g. \citet{sourie17,gavassino20}), these effects are not expected to be so large for slow post-glitch recovery. Nevertheless, GR effects introduce modifications to relaxation time-scale with density dependence. A general relativistic formulation of the vortex creep model is beyond the scope of the present paper and left as a future work.
\section{Glitches in Gamma-Ray Pulsars} \label{sec:glitches}

In this section we present the information on 20 glitches observed from four gamma ray pulsars in our sample and the fits done for 16 large glitches within the framework of the vortex creep model.

\begin{table*}
\caption{Glitches of four gamma-ray pulsars analysed in this work.}
\begin{center}{
\begin{tabular}{lccccccc}
\hline\hline\\
\multicolumn{1}{c}{Pulsar Name} & \multicolumn{1}{c}{Age} & \multicolumn{1}{c}{$B_{\rm s}$} & \multicolumn{1}{c}{Glitch Date} & $\Delta\nu/\nu$ & $\Delta\dot{\nu}/\dot{\nu}$ & Glitch No. & \multicolumn{1}{c}{New or Previous?} \\
& \multicolumn{1}{c}{(kyr)} & \multicolumn{1}{c}{($10^{12}$\,G)} & \multicolumn{1}{c}{(MJD)} & ($10^{-9}$) & ($10^{-3}$) & & \multicolumn{1}{c}{(N/P)}\\ 
\hline\\
J0835$-$4510 & 11.3 & 3.4 & 55408 & 1890(40) & 75(1) & G17 & P\\\\
                        &         &          & 56555 & 3060(40) & 148(1) & G18 & P\\\\
                        &         &          & 57734.4(2) & 1431(2) & 73.3(15) & G20 & P\\\\
                        &         &          & 58515.5929(5) & 2501(2) & 8.6(2) & G21 & P\\\\
J1023$-$5746 & 4.6  & 6.6 & 55043(8) & 3570(1) & 10.62(7) & G1 & P\\\\
                        &       &          & 56004(8) & 2822(4) & 12.6(2) & G2 & N\\\\
                        &       &          & 56647(8) & 3306(1) & 11.0(1) & G3 & N\\\\
                        &       &          & 57192(8) & 22(1) & -1.74(8) & G4 & N\\\\
                        &       &          & 57727(8) & 3784.9(9) & 14.82(2) & G5 & N\\\\
                        &       &          & 58202(17) & 2729(1) & 5.20(3) & G6 & N\\\\
                        &       &          & 58823(8) & 1681.0(4) & 6.77(9) & G7 & N\\\\
J2111$+$4606&17.5 & 4.8  & 55700(5) & 1091.5(1.4) & 3.65(13) & G1 & N\\\\
                         &      &          & 57562(15) & 3258(1) & 7.1(1.7) & G2 & N\\\\
J2229$+$6114 &10.5 & 2.03 & 54777(5) & 0.62(6) & -0.78(8) & G3 & P\\\\
                         &      &          & 55134(1) & 195(2) & 5.5(1) & G4 & P\\\\
                         &      &          & 55601(2) & 1222(4) & 12.90(1) & G5 & P\\\\
                         &      &          & 56366.59(2) & 65.9(2) & 3.32(2) & G6 & P\\\\
                         &      &          & 56797(5) & 850(1) & 11.3(1) & G7 & N\\\\
                         &      &          & 58425(2) & 1032.8(2)& 11.8(1)& G8& N$^*$ \\\\
                         &      &          & 59270(2)& 277.2(1)& 7.38(9)& G9& N\\\\
\hline\\
\label{glitchobs}
\end{tabular}}
\begin{tablenotes}

Note~$^*$--\citet{basu21} later obtained the following parameters for G8 of PSR J2229+6114 which are consistent with our results: Glitch Date ( MJD) 58424(6), $\Delta\nu/\nu~(10^{-9})=1047.1(9)$, and  $\Delta\dot\nu/\dot\nu~(10^{-3})=12.4(6)$.

\end{tablenotes}
\end{center}
\end{table*}

\subsection{PSR J0835$-$4510 (Vela)} \label{sec:PSR0835}

The Vela pulsar is one of the most prolific glitchers with 18 large glitches occurred quasi-periodically in every 2-3 years since its discovery in 1967. We analysed the Vela pulsar's timing data between MJD 54692-58839. In this period Vela glitched four times with all of its glitches are large ones and glitch activity parameter is $A_{\rm g}=7.9\times10^{-7}$\yr. These four glitches, which are clearly seen in Fig.\ref{J0835spin}, were also detected previously in the literature except for its last glitch for which the post-glitch recovery data were not published before. Glitch dates and magnitudes are listed in Table \ref{glitchobs}.

The 2010 and 2013 glitches of the Vela pulsar were also assessed with the vortex creep model by \citet{akbal17}. However, \citet{akbal17} essentially concentrated on the long term post-glitch behaviour and used the following function in their fits:
\begin{equation}
\Delta\dot\nu(t)=-\frac{I_{\ell}}{I}\frac{\Delta\nu}{[\tau_{\ell}=30~\mbox{days}]}\exp\left(-\frac{t}{30}\right)+\frac{I_{\rm A}}{I}\left(1-\frac{t}{t_{0}}\right)\dot\nu_{0}\nonumber,
\end{equation}
by assigning exponential decay time-scale $\tau_{\ell}=30$ days, a representative value for the Vela. Here we use the full solution, i.e. Eqs.(\ref{creepexp}), (\ref{creepfull}), and leave all fit parameters free in order to analyse the full data better and extract information on the various layers contributing both to short term exponential recovery and long-term relaxation. Thus, our fit results slightly differ from \citet{akbal17}.

\begin{figure}
\centering
\vspace{0.1cm}
\includegraphics[width=1.0\linewidth]{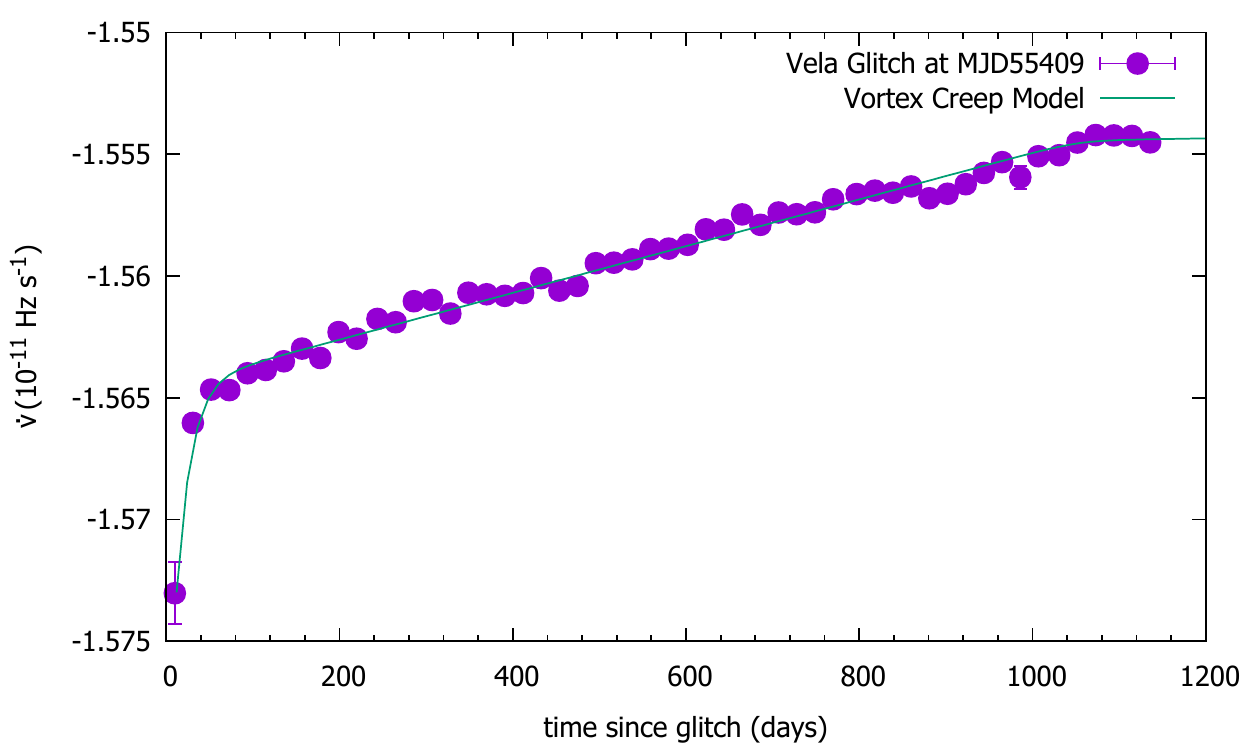}
\caption{Comparison between the 2010 Vela post-glitch data (purple) and the vortex creep model (green).}
\label{vela2010}
\end{figure}

We fit the 2010 Vela glitch data with Eqs.(\ref{creepexp}), (\ref{creepfull}) and the fit result is shown in Fig. \ref{vela2010}. Fit parameters are given in Table \ref{modelfit}. The vortex creep model prediction for the time to the next glitch from the 2010 glitch is $t_{\rm th}=984(16)$ days and as can be seen from Fig. \ref{vela2010} recovery of this glitch should have already been completed 4 data points before the occurrence of the 2013 glitch (i.e. around MJD 56460), which is in close agreement with our theoretical estimation. The total fractional crustal superfluid moment of inertia participated in the 2010 Vela glitch is $I_{\rm cs}/I=1.95(4)\times10^{-2}$.

\begin{figure}
\centering
\vspace{0.1cm}
\includegraphics[width=1.0\linewidth]{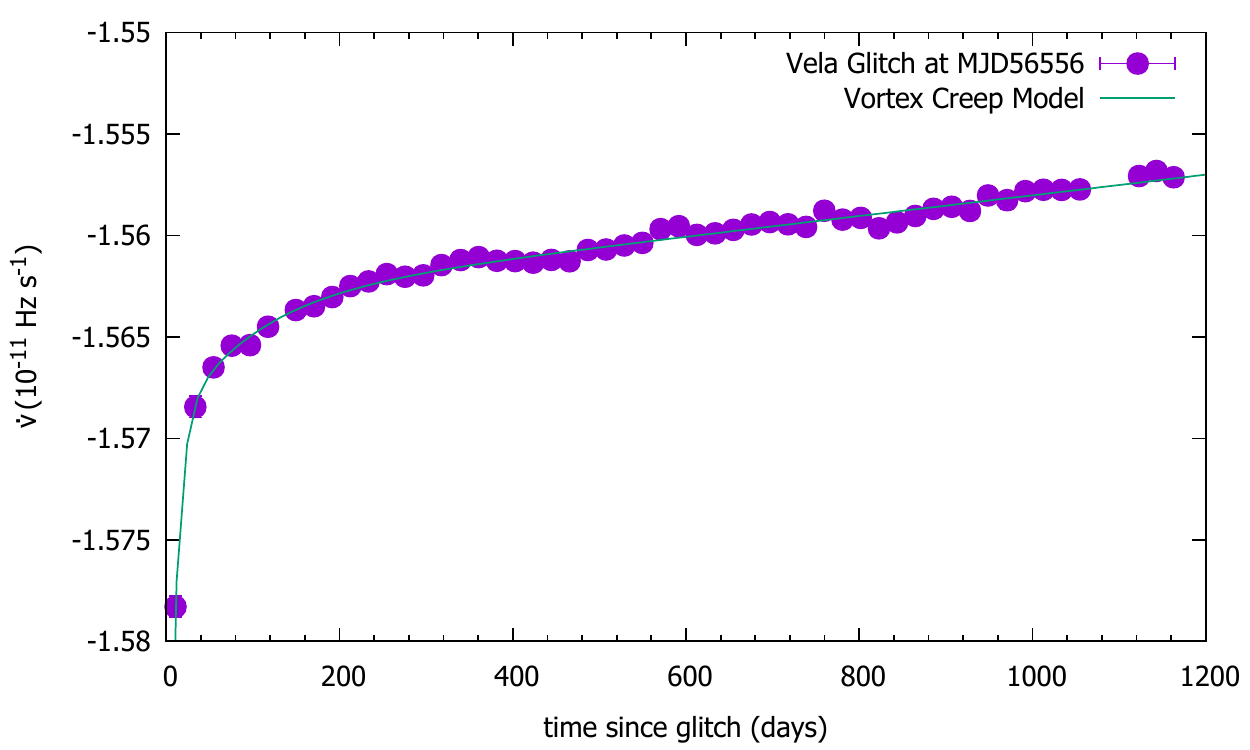}
\caption{Comparison between the 2013 Vela post-glitch data (purple) and the vortex creep model (green).}
\label{vela2013}
\end{figure}

We fit the 2013 Vela glitch with Eqs.(\ref{creepexp}), (\ref{glitchdotnu}) and the fit result is shown in Fig. \ref{vela2013}. Fit parameters are given in Table \ref{modelfit}. The vortex creep model prediction for the time to the next glitch from the 2013 glitch is $t_{\rm th}=1862(48)$ days and the observed time-scale is $t_{\rm obs}=1178$ days. But as can be seen from Fig. \ref{J0835spin}, the last post-glitch data point of the 2013 Vela glitch is clearly below the general trend so that the 2016 glitch arrived at before the recovery of the 2013 glitch had been fully completed. The total fractional crustal superfluid moment of inertia involved in the 2013 glitch is $I_{\rm cs}/I=1.77(6)\times10^{-2}$. 

\begin{figure}
\centering
\vspace{0.1cm}
\includegraphics[width=1.0\linewidth]{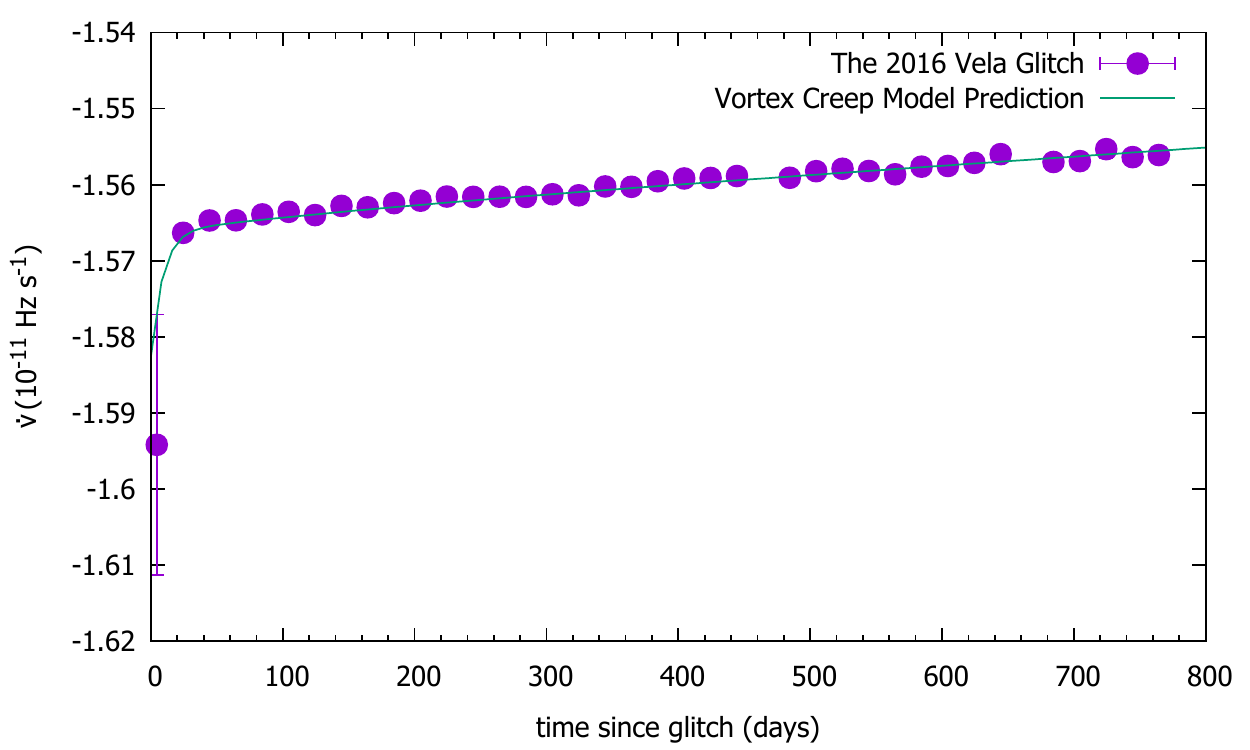}
\caption{Comparison between the 2016 Vela post-glitch data (purple) and the vortex creep model (green).}
\label{vela2016}
\end{figure}

The first 416 days post-glitch timing data of the 2016 Vela glitch were fitted with Eqs.(\ref{creepexp}), (\ref{creepfull}) by \citet{erbil20}. Here we fit 765 days long full data of the 2016 Vela glitch with the same set of parameters evaluated by \citet{erbil20} and the resulting fit is shown in Fig. \ref{vela2016}. The vortex creep model prediction for the time to the next glitch from the 2016 glitch is $t_{\rm th}=781(13)$ days which is in excellent agreement with the observed time-scale $t_{\rm obs}=782$ days. \citet{erbil20} constrained the vortex line-flux tube pinning energy as 2 MeV and determined the critical lag for vortex unpinning as $\omega_{\rm cr}=0.01$\rads. The total fractional crustal superfluid moment of inertia participated in the 2016 glitch is $I_{\rm cs}/I=1.68(4)\times10^{-2}$. 

\begin{figure}
\centering
\vspace{0.1cm}
\includegraphics[width=1.0\linewidth]{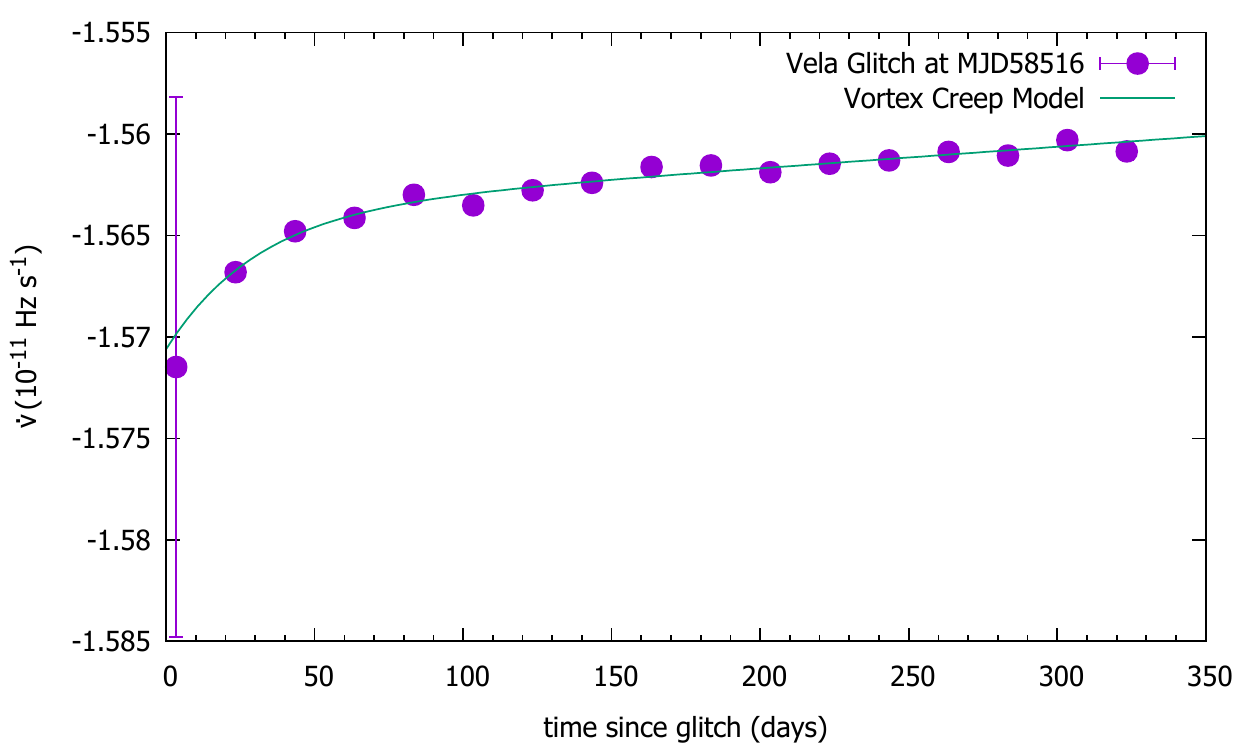}
\caption{Comparison between the 2019 Vela post-glitch data (purple) and the vortex creep model (green).}
\label{vela2019}
\end{figure}

For the 2019 Vela glitch \citet{lower20} only reported on the glitch magnitudes $\Delta\nu/\nu=2.5\times10^{-6}$ and $\Delta\dot\nu/\dot\nu=9\times10^{-3}$ but did not discuss the post-glitch evolution. Here we, for the first time, present the post-glitch recovery of the 2019 glitch of the Vela pulsar starting from 4 days after the occurrence of the glitch and extending to 325 days afterwards which is shown in Fig. \ref{vela2019}. The vortex creep model prediction for the time to the next glitch from the parameters of the 2019 Vela glitch is $t_{\rm th}=769(88)$ days \footnote{The next large glitch of the Vela pulsar occurred on 2021 July 23 (MJD 59418) \citep{sosa21} with a 45 days deviation from our expectation. A small size event with $\Delta\nu/\nu\gtrsim10^{-9}$ intervening the recovery of the 2019 glitch may have been taken place which affected the excess superfluid angular momentum accumulation rate and caused the delay.}. The total fractional crustal superfluid moment of inertia involved in the 2019 Vela glitch is $I_{\rm cs}/I=2.94(28)\times10^{-2}$.

\subsection{PSR J1023$-$5746} \label{sec:PSR1023}

We analysed the timing data of PSR J1023$-$5746 between MJD 54728-59269. In this period PSR J1023$-$5746 glitched 7 times with 6 of them are large glitches and glitch activity parameter is $A_{\rm g}=14.5\times10^{-7}$\yr. Of these, only the first glitch was observed previously by \citet{sazparkinson10} and the remaining glitches are new. Glitch dates and magnitudes are listed in Table \ref{glitchobs}. Six large glitches of PSR J1023$-$5746 are clearly seen in Fig.\ref{J1023spin}.

\begin{figure}
\centering
\vspace{0.1cm}
\includegraphics[width=1.0\linewidth]{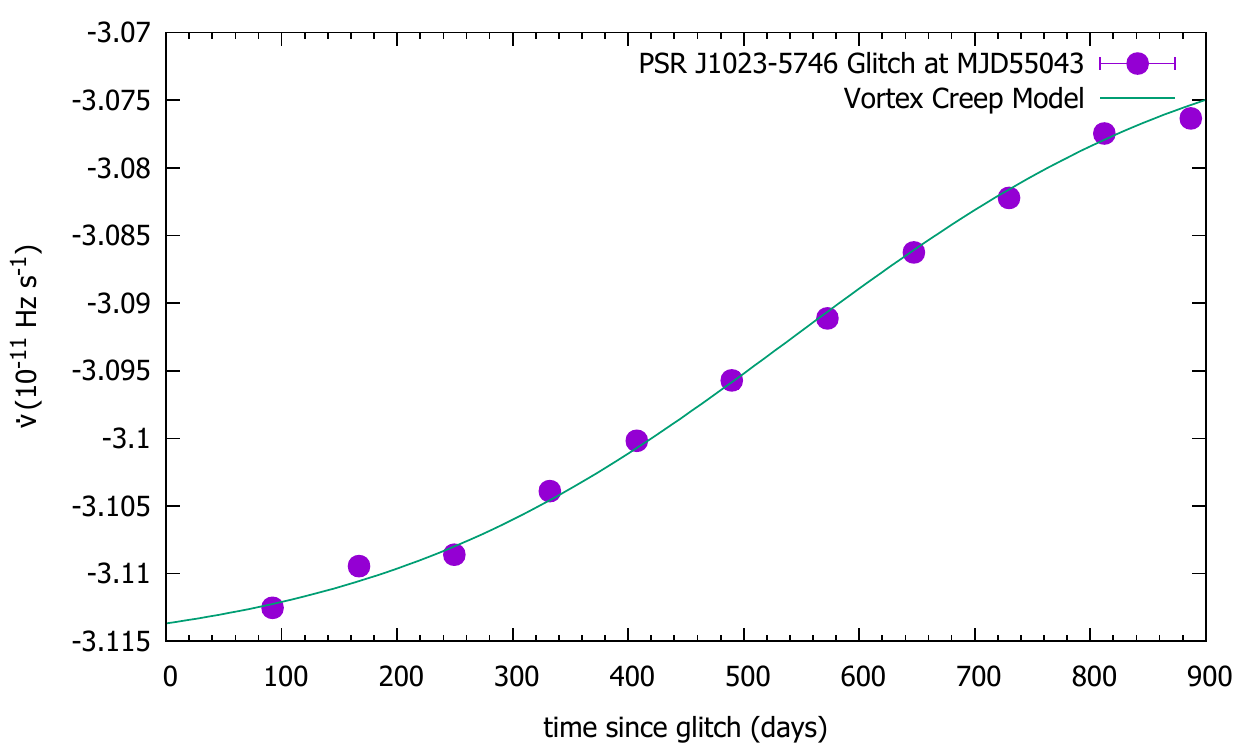}
\caption{Comparison between the post-glitch data of PSR J1023$-$5746 for the glitch occurred at MJD55043 (purple) and the vortex creep model (green).}
\label{J1023G1}
\end{figure}

We fit the first glitch (G1) of PSR J1023$-$5746 with Eq.(\ref{creepsingle}) and the fit result is shown in Fig. \ref{J1023G1}. Fit parameters are given in Table \ref{modelfit}. The vortex creep model prediction for the time to the next glitch from G1 is $t_{\rm th}=1110(31)$ days, in agreement with the observed time-scale $t_{\rm obs}=961(16)$ days. The total fractional crustal superfluid moment of inertia participated in G1 is $I_{\rm cs}/I=2.19(7)\times10^{-2}$.

\begin{figure}
\centering
\vspace{0.1cm}
\includegraphics[width=1.0\linewidth]{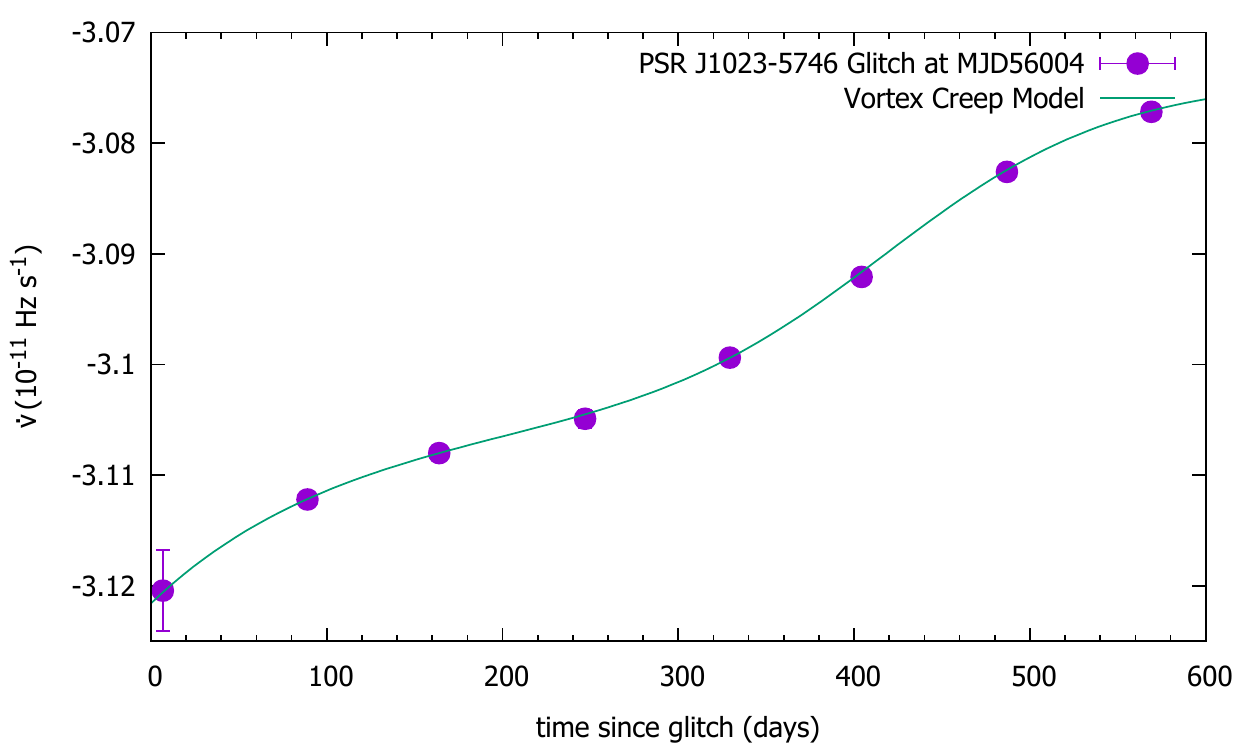}
\caption{Comparison between the post-glitch data of PSR J1023$-$5746 for the glitch occurred at MJD56004 (purple) and the vortex creep model (green).}
\label{J1023G2}
\end{figure}

We fit the second glitch (G2) of PSR J1023$-$5746 with Eqs.(\ref{creepexp}), (\ref{creepsingle}) and the fit result is shown in Fig. \ref{J1023G2}. Fit parameters are given in Table \ref{modelfit}. The vortex creep model prediction for the time to the next glitch from G2 is $t_{\rm th}=620(15)$ days, in close agreement with the observed time-scale $t_{\rm obs}=643(16)$ days. The total fractional crustal superfluid moment of inertia involved in G2 is $I_{\rm cs}/I=2.26(10)\times10^{-2}$.

\begin{figure}
\centering
\vspace{0.1cm}
\includegraphics[width=1.0\linewidth]{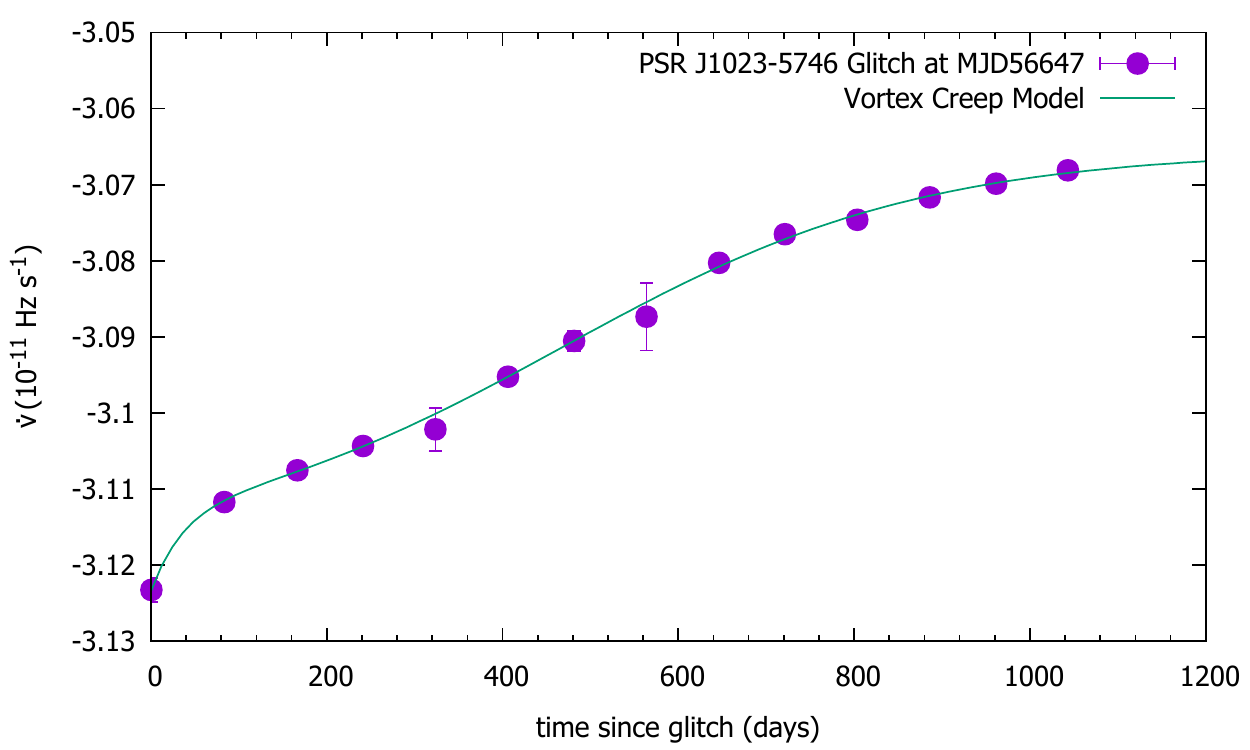}
\caption{Comparison between the post-glitch data of PSR J1023$-$5746 for the glitch occurred at MJD56647 (purple) and the vortex creep model (green).}
\label{J1023G3}
\end{figure}

We fit the third glitch (G3) of PSR J1023$-$5746 with Eqs.(\ref{creepexp}), (\ref{creepsingle}) and the fit result is shown in Fig. \ref{J1023G3}. Fit parameters are given in Table \ref{modelfit}. The vortex creep model prediction for the time to the next glitch from G3 is $t_{\rm th}=1102(47)$ days which matches with the observed $t_{\rm obs}=1080(16)$ days. The total fractional crustal superfluid moment of inertia participated in G3 is $I_{\rm cs}/I=2.30(16)\times10^{-2}$.

The fourth glitch (G4) of PSR J1023$-$5746 is quite small and does not lead to any appreciable change in the spin-down rate. So, we ignore this glitch in our vortex creep model fit analyses.

\begin{figure}
\centering
\vspace{0.1cm}
\includegraphics[width=1.0\linewidth]{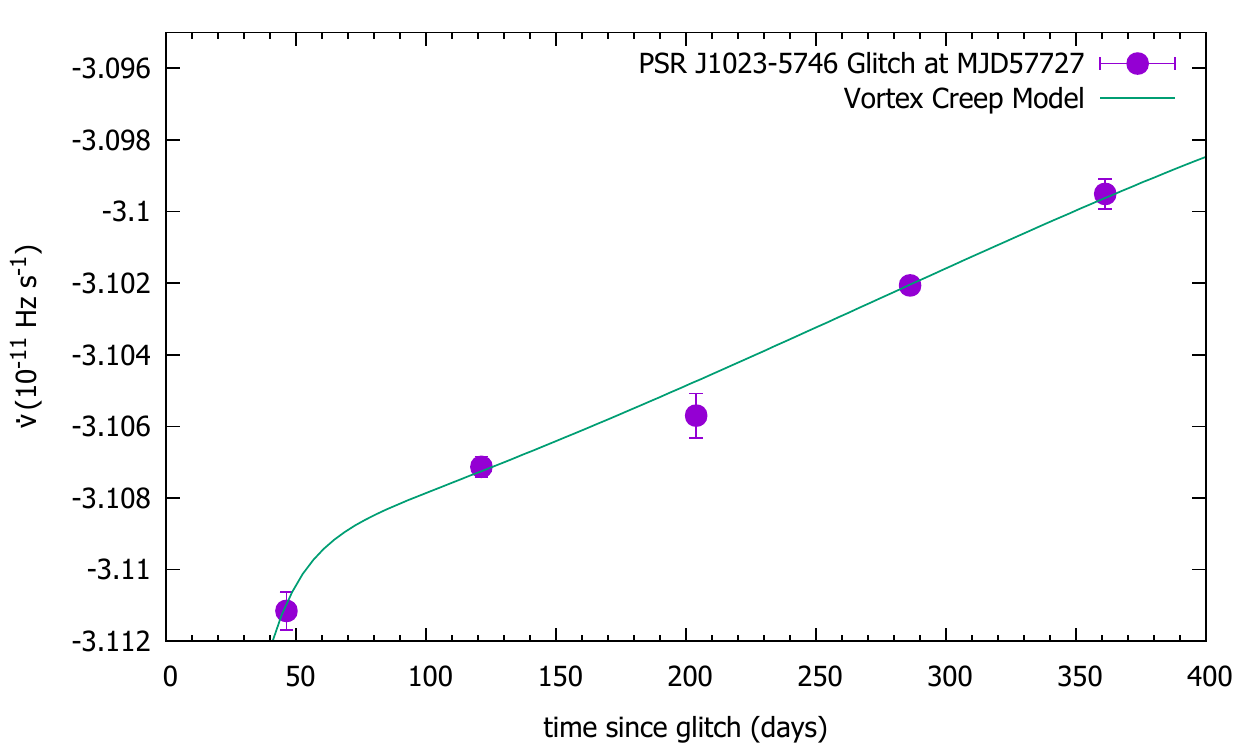}
\caption{Comparison between the post-glitch data of PSR J1023$-$5746 for the glitch occurred at MJD57727 (purple) and the vortex creep model (green).}
\label{J1023G5}
\end{figure}

We fit the fifth glitch (G5) of PSR J1023$-$5746 with Eqs.(\ref{creepexp}), (\ref{creepfull}) and the fit result is shown in Fig. \ref{J1023G5}. Fit parameters are given in Table \ref{modelfit}. The vortex creep model prediction for the time to the next glitch from G5 is $t_{\rm th}=484$ days and has excellent agreement with the observed $t_{\rm obs}=475(25)$ days. The total fractional crustal superfluid moment of inertia involved in G5 is $I_{\rm cs}/I=2.89\times10^{-2}$.

\begin{figure}
\centering
\vspace{0.1cm}
\includegraphics[width=1.0\linewidth]{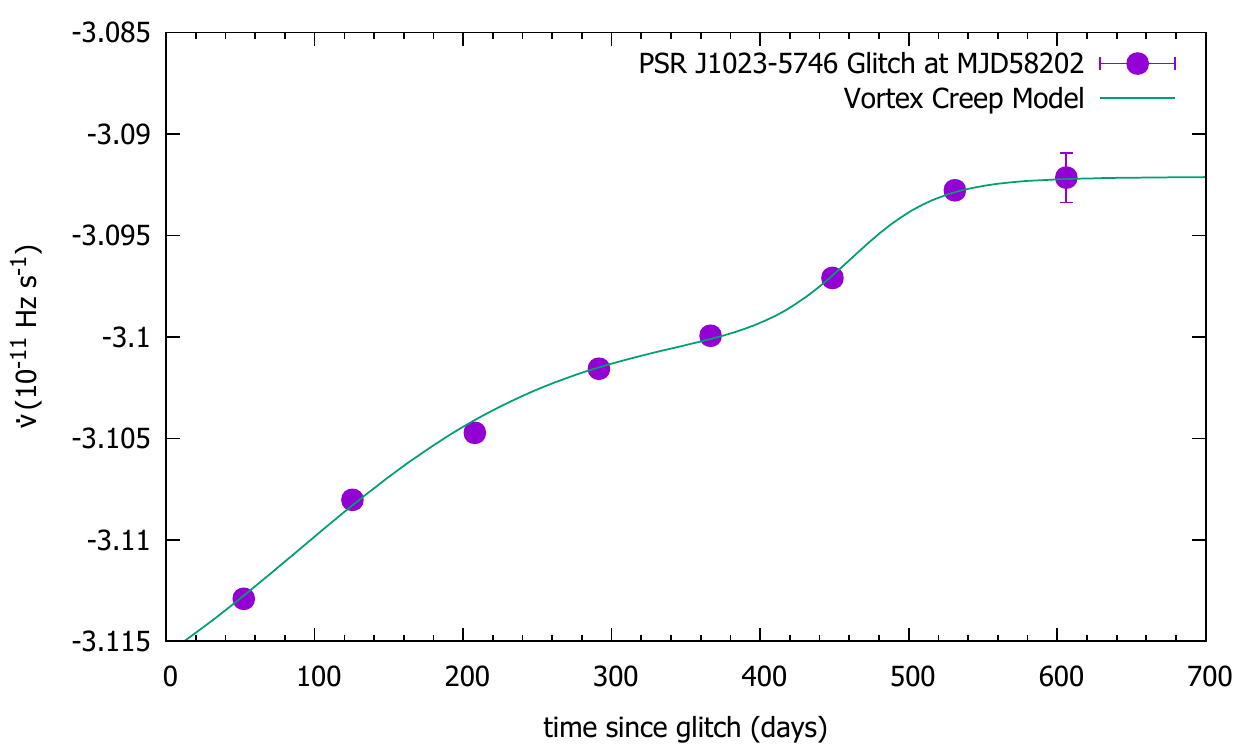}
\caption{Comparison between the post-glitch data of PSR J1023$-$5746 for the glitch occurred at MJD58202 (purple) and the vortex creep model (green).}
\label{J1023G6}
\end{figure}

We fit the sixth glitch (G6) of PSR J1023$-$5746 with Eq.(\ref{creepsingle}) and the fit result is shown in Fig. \ref{J1023G6}. Fit parameters are given in Table \ref{modelfit}. The vortex creep model prediction for the time to the next glitch from G6 is $t_{\rm th}=548(5)$ days while the observed time-scale is $t_{\rm obs}=621(25)$ days. The total fractional crustal superfluid moment of inertia participated in G6 is $I_{\rm cs}/I=2.2(2)\times10^{-2}$.

\begin{figure}
\centering
\vspace{0.1cm}
\includegraphics[width=1.0\linewidth]{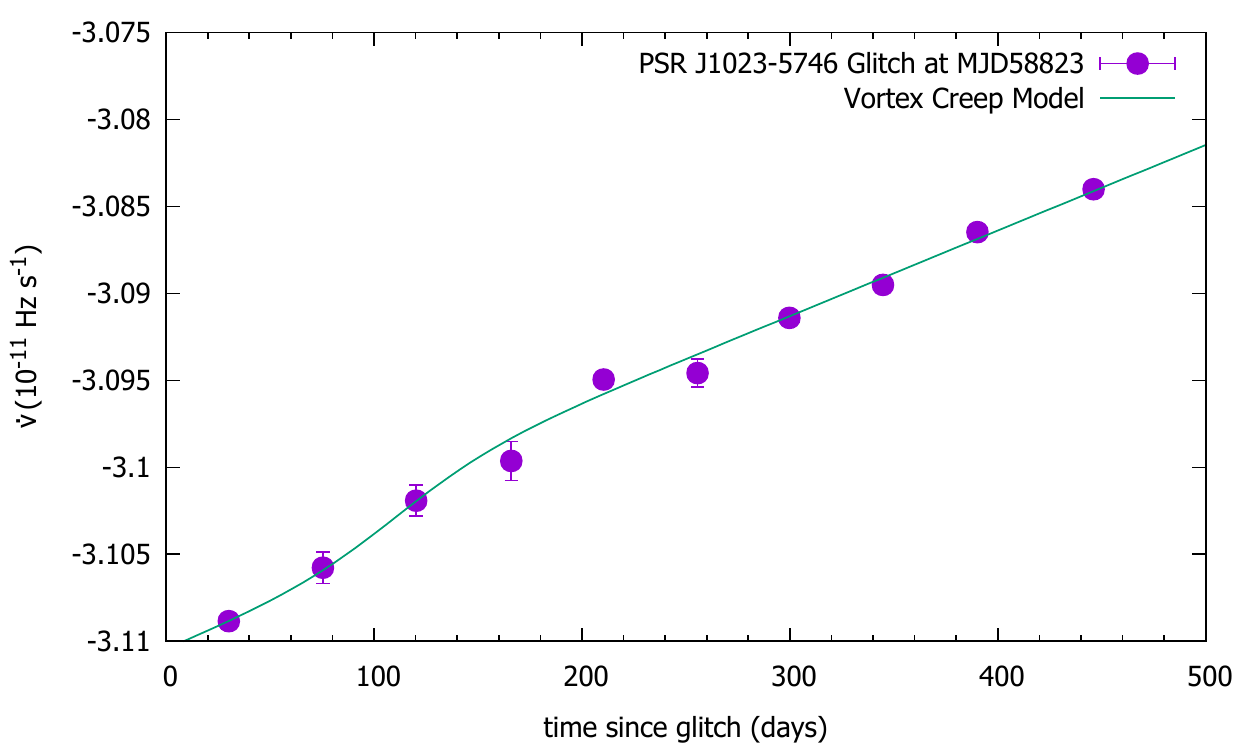}
\caption{Comparison between the post-glitch data of PSR J1023$-$5746 for the glitch occurred at MJD58823 (purple) and the vortex creep model (green).}
\label{J1023G7}
\end{figure}

We fit the seventh glitch (G7) of PSR J1023$-$5746 with Eqs.(\ref{creepsingle}), (\ref{glitchdotnu}) and the fit result is shown in Fig. \ref{J1023G7}. Fit parameters are given in Table \ref{modelfit}. The vortex creep model prediction for the time to the next glitch from G7 is $t_{\rm th}=862(97)$ days and the total fractional crustal superfluid moment of inertia participated in G7 is $I_{\rm cs}/I=0.77(10)\times10^{-2}$. Both the magnitude and fractional moment of inertia of the crustal superfluid regions involved in G7 are smaller than typical values for previous PSR J1023--5746 glitches. The underlying reason is that G7 arrived at earlier than the recovery of G6 had been fully completed, as is evident from Fig.\ref{J1023spin}. Therefore instability giving rise to the glitch did not cover the whole crust for G7 case. Due to the long waiting time of G7, we expect that next glitch of the PSR J1023$-$5746 will be a larger one, typical for this pulsar.

An inspection of Table \ref{modelfit} reveals that for PSR J1023$-$5746 the superfluid recoupling time-scales $\tau_{\rm s}$, $\tau_{\rm nl}$, the offset times $t_{\rm s}$, $t_{0}$ which are related to the number of vortex lines participated in the glitches, and also the total crustal moment of inertia $I_{\rm cs}$ involved in its glitches do not change from one glitch to the other significantly. This suggests that for PSR J1023$-$5746 same crustal superfluid regions are taking part in its glitches.

\subsection{PSR J2111$+$4606} \label{sec:PSR2111}

We analysed the timing data of PSR J2111$+$4606 between MJD 54787-58598.  In this period PSR J2111$+$4606 glitched twice with both of its glitches are large ones and glitch activity parameter is $A_{\rm g}=4.2\times10^{-7}$\yr. These two glitches, clearly seen in Fig.\ref{J2111spin}, are new discoveries. Glitch dates and magnitudes are listed in Table \ref{glitchobs}.

\begin{figure}
\centering
\vspace{0.1cm}
\includegraphics[width=1.0\linewidth]{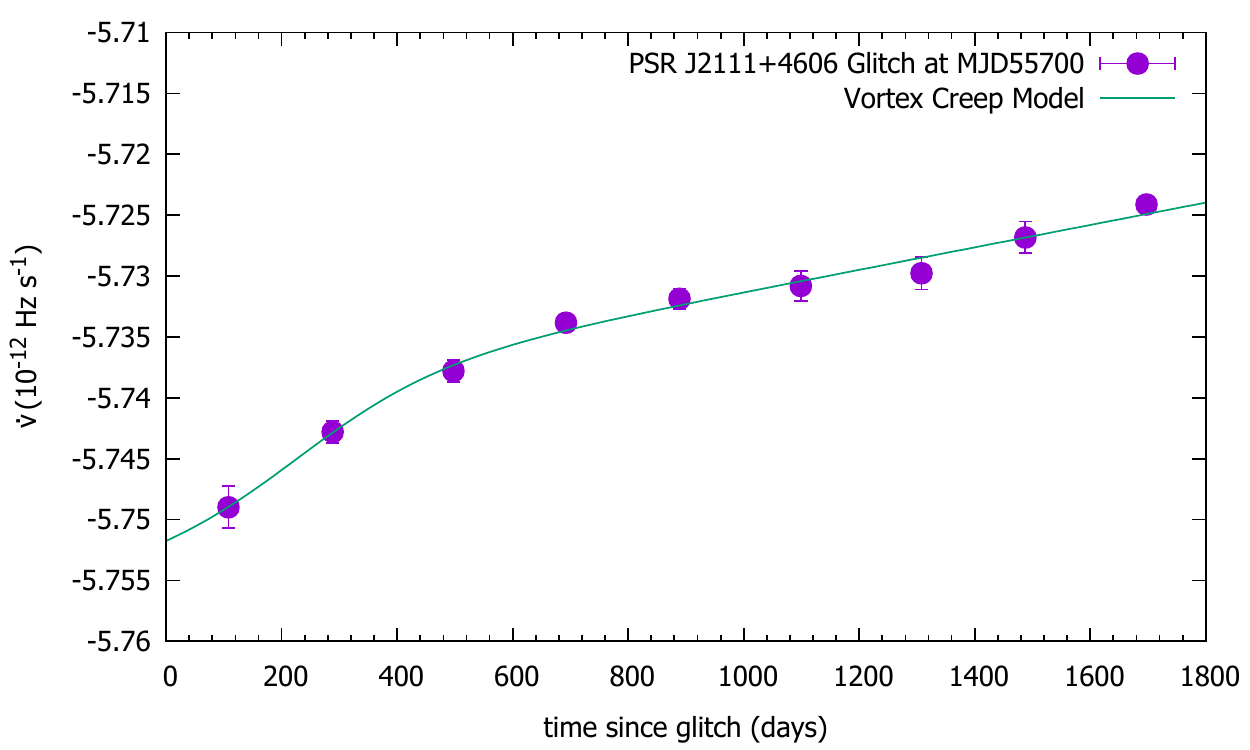}
\caption{Comparison between the post-glitch data of PSR J2111$+$4606 for the glitch occurred at MJD55700 (purple) and the vortex creep model (green). }
\label{J2111G1}
\end{figure}

We fit the first glitch (G1) of PSR J2111$+$4606 with Eqs.(\ref{creepsingle}), (\ref{glitchdotnu}) and the fit result is shown in Fig. \ref{J2111G1}. Fit parameters are given in Table \ref{modelfit}. The vortex creep model prediction for the time to the next glitch from G1 is $t_{\rm th}=1864(90)$ days and has excellent agreement with the observed inter-glitch time-scale $t_{\rm obs}=1862(20)$ days. The total fractional crustal superfluid moment of inertia participated in G1 is $I_{\rm cs}/I=1.03(20)\times10^{-2}$ and it does not cover the whole crust. This is consistent with the fact that the vortex unpinning avalanche started at moderate depths of the crust.

\begin{figure}
\centering
\vspace{0.1cm}
\includegraphics[width=1.0\linewidth]{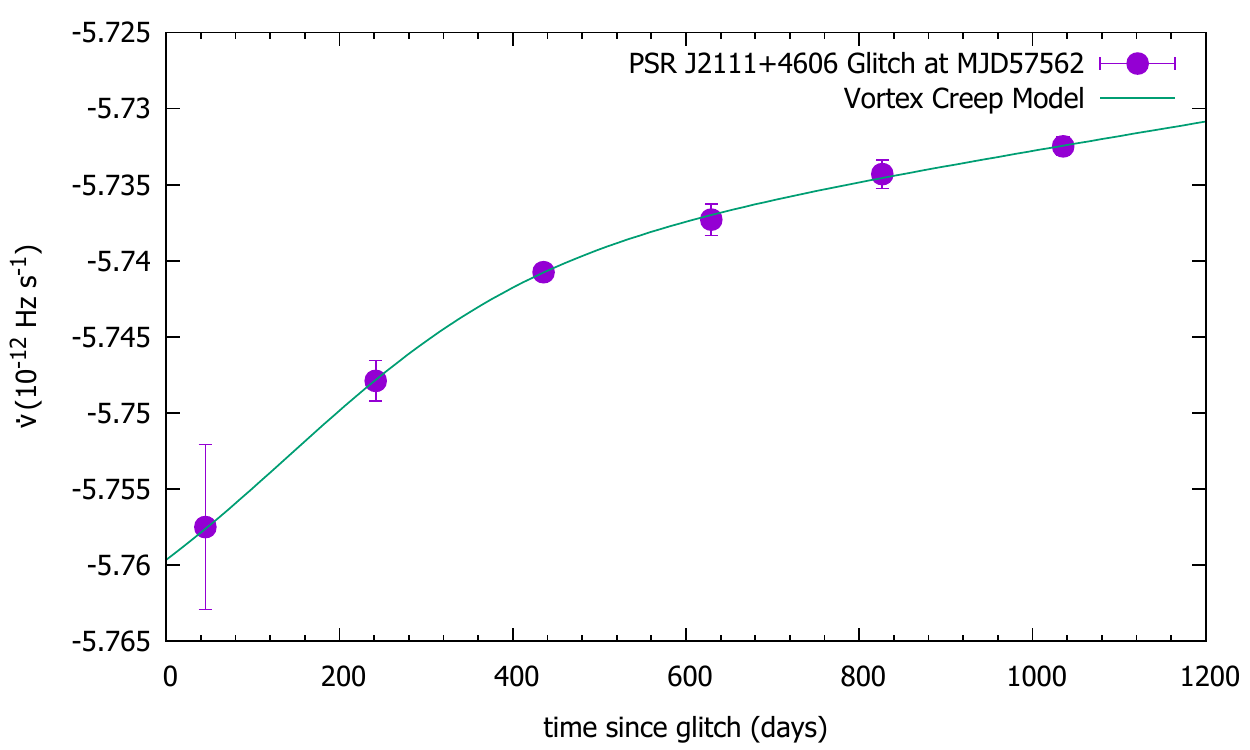}
\caption{Comparison between the post-glitch data of PSR J2111$+$4606 for the glitch occurred at MJD57562 (purple) and the vortex creep model (green).}
\label{J2111G2}
\end{figure}

We fit the second glitch (G2) of PSR J2111$+$4606 with Eqs.(\ref{creepsingle}), (\ref{glitchdotnu}) and the fit result is shown in Fig. \ref{J2111G2}. Fit parameters are given in Table \ref{modelfit}. The vortex creep model prediction for the time to the next glitch from G1 is $t_{\rm th}=2140(311)$ days and we expect that the next large glitch of PSR J2111$+$4606 will not come earlier than mid June 2021, and likely to occur around 2022 May 3. The total fractional crustal superfluid moment of inertia participated in G2 is $I_{\rm cs}/I=2.37(34)\times10^{-2}$. Both the glitch magnitude and the implied total fractional crustal superfluid moment of inertia for G2 are larger compared to G1 case. The reason is involvement of larger capacitive superfluid region $I_{\rm B}$ in G2. 

Superfluid recoupling time-scales $\tau_{\rm s}$, $\tau_{\rm nl}$ and offset times $t_{\rm s}$, $t_{0}$ do not deviate significantly among G1 and G2 of PSR J2111$+$4606, suggesting that the instability leading to its glitches occur at the same location for this pulsar and the main parameter determining the final magnitude of its glitches is the size of the vortex depletion region $I_{\rm B}$ that unpinned vortices move through.

\subsection{PSR J2229$+$6114} \label{sec:PSR2229}

We analysed the timing data of PSR J2229$+$6114 between MJD 54705-59473. In this period PSR J2229$+$6114 glitched seven times with three of them are large glitches and glitch activity parameter is $A_{\rm g}=2.8\times10^{-7}$\yr. Among these glitches first four events were observed previously \citep{espinoza11} and the remaining three are new\footnote{Later, \citet{basu21} also observed G8 of PSR J2229+6114  in radio wavelengths. For consistency their glitch parameters are given in the note of Table \ref{glitchobs}.}. Glitch dates and magnitudes are listed in Table \ref{glitchobs}. Six glitches of PSR J2229$+$6114 with $\Delta\dot\nu/\dot\nu\gtrsim10^{-3}$ can be clearly seen in Fig.\ref{J2229spin}.

The third glitch (G3) of PSR J2229$+$6114 is small and do not lead to appreciable changes in the general spin-down rate behaviour. So we ignore this glitch in our vortex creep model analyses.

\begin{figure}
\centering
\vspace{0.1cm}
\includegraphics[width=1.0\linewidth]{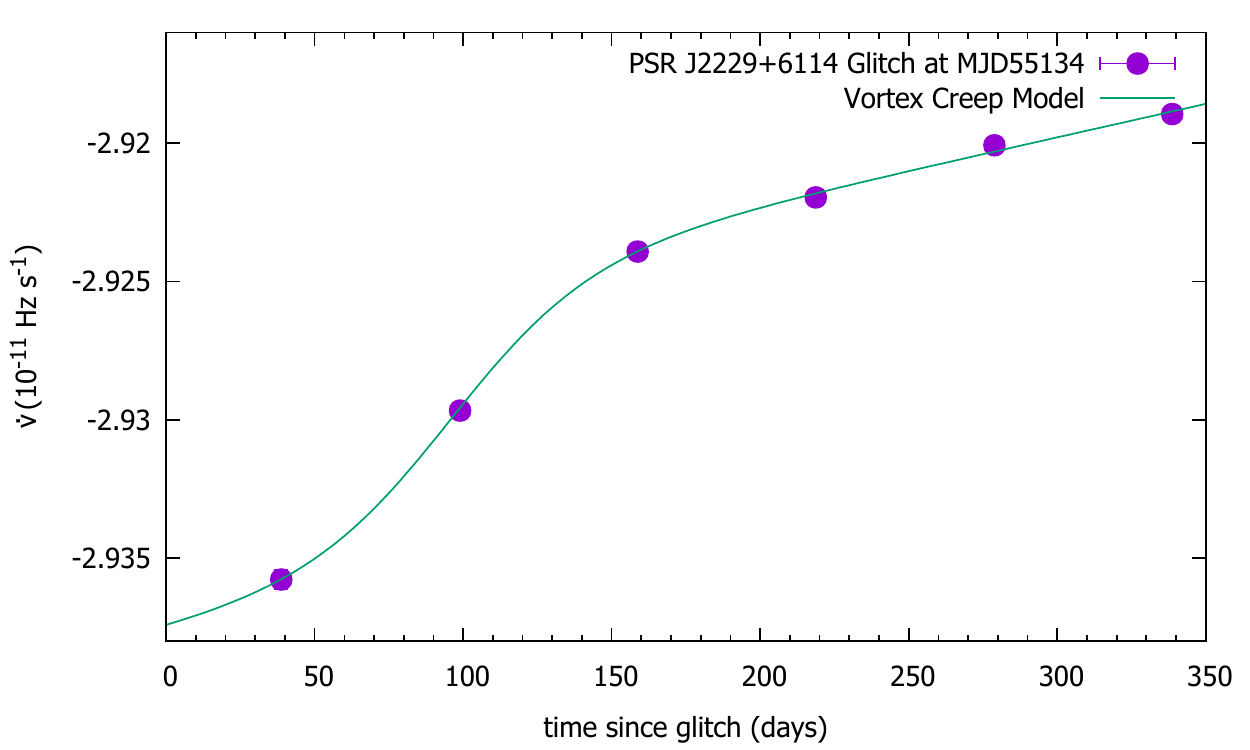}
\caption{Comparison between the post-glitch data of PSR J2229$+$6114 for the glitch occurred at MJD55134 (purple) and the vortex creep model (green).}
\label{J2229G4}
\end{figure}

We fit the fourth glitch (G4) of PSR J2229$+$6114 with Eqs.(\ref{creepsingle}), (\ref{glitchdotnu}) and the fit result is shown in Fig. \ref{J2229G4}. Fit parameters are given in Table \ref{modelfit}. The vortex creep model prediction for the time to the next glitch from G4 is $t_{\rm th}=452(28)$ days which is in excellent agreement with observed $t_{\rm obs}=460(12)$ days. The total fractional crustal superfluid moment of inertia involved in G4 is $I_{\rm cs}/I=0.79(9)\times10^{-2}$.

\begin{figure}
\centering
\vspace{0.1cm}
\includegraphics[width=1.0\linewidth]{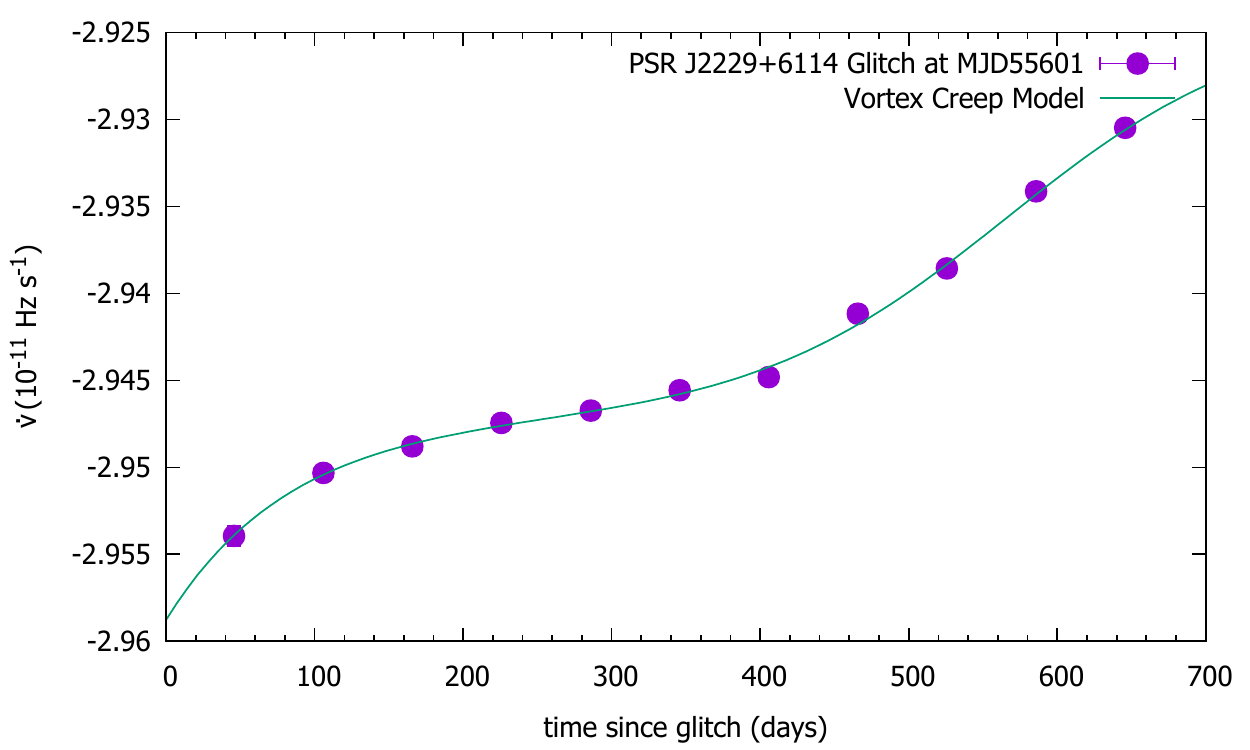}
\caption{Comparison between the post-glitch data of PSR J2229$+$6114 for the glitch occurred at MJD55601 (purple) and the vortex creep model (green).}
\label{J2229G5}
\end{figure}

We fit the fifth glitch (G5) of PSR J2229$+$6114 with Eqs.(\ref{creepexp}), (\ref{creepsingle}) and the fit result is shown in Fig. \ref{J2229G5}. Fit parameters are given in Table \ref{modelfit}. The vortex creep model prediction for the time to the next glitch from G5 is $t_{\rm th}=821(61)$ days, in agreement with observed $t_{\rm obs}=760(10)$ days within errors. The total fractional crustal superfluid moment of inertia involved in G5 is $I_{\rm cs}/I=1.66(22)\times10^{-2}$.

The sixth glitch (G6) of PSR J2229$+$6114 is a small event between two large glitches with no sizable effect on the spin-down rate trend. So we ignore this glitch from our fit analyses.

\begin{figure}
\centering
\vspace{0.1cm}
\includegraphics[width=1.0\linewidth]{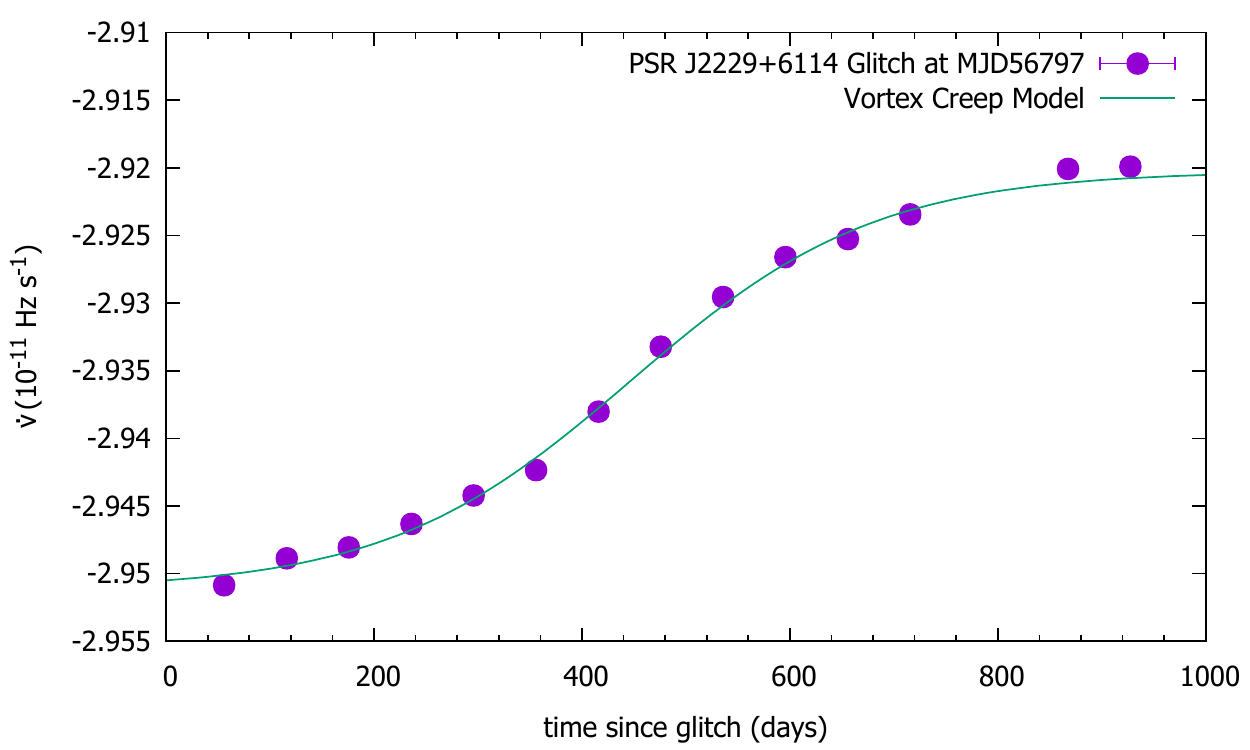}
\caption{Comparison between the post-glitch data of PSR J2229$+$6114 for the glitch occurred at MJD56797 (purple) and the vortex creep model (green).}
\label{J2229G7}
\end{figure}

We fit the seventh glitch (G7) of PSR J2229$+$6114 with Eq.(\ref{creepsingle}) and the fit result is shown in Fig. \ref{J2229G7}. Fit parameters are given in Table \ref{modelfit}. The total fractional crustal superfluid moment of inertia participated in G7 is $I_{\rm cs}/I=1.44(4)\times10^{-2}$. The vortex creep model prediction for the time to the next glitch from G7 is $t_{\rm th}=795(17)$ days while the observed time-scale is $t_{\rm obs}=975(8)$ days. According to the vortex creep model, after the offset time $t_{0}$  has elapsed a pulsar resumes to its steady spin-down under true braking mechanism so that the glitch contribution to the rotational evolution ceases. Then, we are able to find $\ddot\nu=1.16(13)\times10^{-22}$\hzss,~which is purely pulsar braking torque origin. From this value, we determined a braking index n=2.63(30) for PSR J2229$+$6114 after glitch induced contributions have been removed. 

\begin{figure}
\centering
\vspace{0.1cm}
\includegraphics[width=1.0\linewidth]{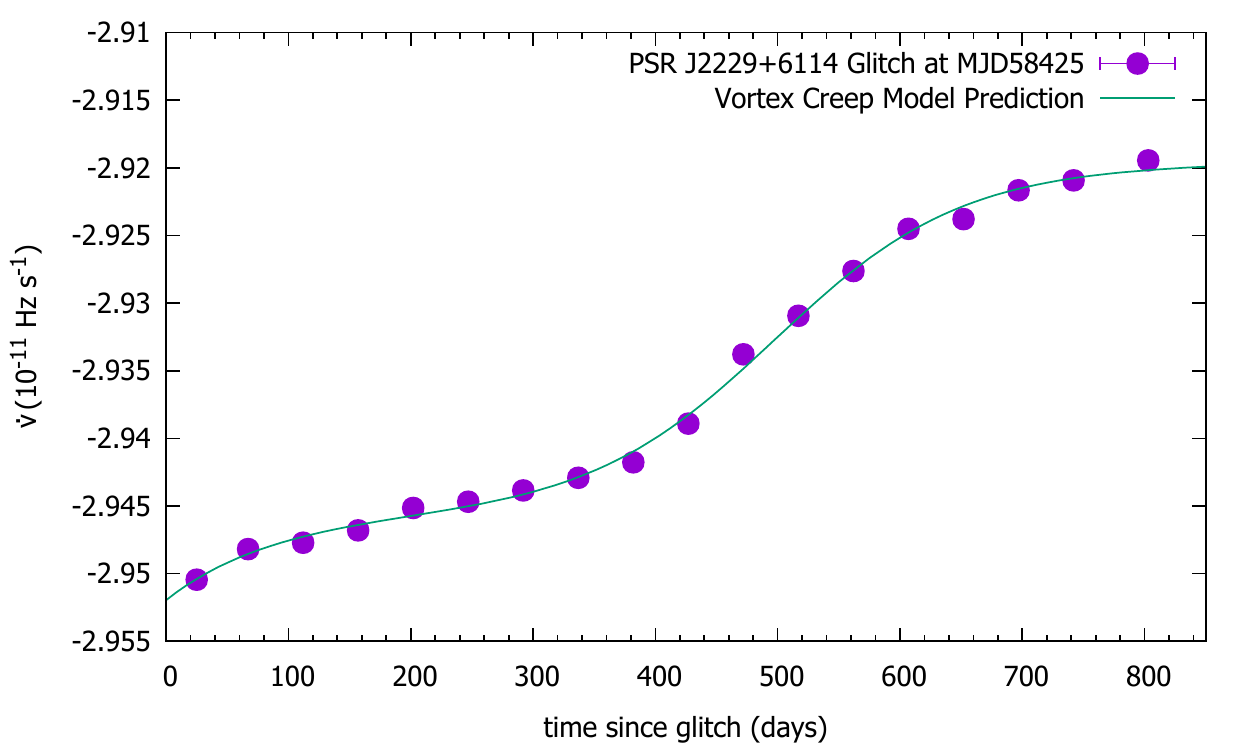}
\caption{Comparison between the post-glitch data of PSR J2229$+$6114 for the glitch occurred at MJD58425 (purple) and the vortex creep model (green).}
\label{J2229G8}
\end{figure}

We fit the eighth glitch (G8) of PSR J2229$+$6114 with Eqs.(\ref{creepexp}), (\ref{creepsingle}) and the fit result is shown in Fig. \ref{J2229G8}. Fit parameters are given in Table \ref{modelfit}. The vortex creep model prediction for the time to the next glitch from G8 is $t_{\rm th}=857(27)$ days, in excellent agreement with the observed time-scale $t_{\rm obs}=845(4)$ days within errors. The total fractional crustal superfluid moment of inertia involved in G8 is $I_{\rm cs}/I=1.62(6)\times10^{-2}$.

\begin{table*}
\caption{Vortex creep model fit results for 16 large glitches in four gamma-ray pulsars analysed in this work.}
\begin{center}{\scriptsize
\begin{tabular}{lccccccccccccc}
\hline\hline\\
\multicolumn{1}{c}{Pulsar Name} & \multicolumn{1}{c}{Glitch Date} & $I_{\rm exp}/I$ & $\tau_{\rm exp}$ & $I_{\rm A}/I$ & $\tau_{\rm nl}$ & $t_{0}$ & $I_{\rm s}/I$ &  $\tau_{\rm s}$ &  $t_{\rm s}$ & $I_{\rm B}/I$ & $t_{\rm th}$ & $t_{\rm obs}$ &  $I_{\rm cs}/I$ \\
& \multicolumn{1}{c}{(MJD)} & ($10^{-3}$) & (days) & ($10^{-3}$) & (days) & (days) & ($10^{-3}$) & (days) & (days) & ($10^{-3}$) & (days) & (days) &  ($10^{-2}$) \\ 
\hline\\
J0835$-$4510 & 55408 & 8.76(3.64) & 17(4) & 6.12(8) &19(34) & 984(16) & - & - & - & 13.3(4) & 984(16) & 1147 & 1.95(4)\\\\
                        & 56555 & 9.26(2.21) & 15(3) & 5.63(9) & - & 1862(48) & 1.95(55) & 37(55) & 146(24) & 10.1(3) & 1862(48) & 1178 & 1.77(6)\\\\
                        & 57734.4(2) & 18.2(4.9) & 9(1) & 3.48(21) & 45(10) & 781(13) & - & - & - & 13.3(5) & 781(13) & 782 & 1.68(4)\\\\
                        & 58515.5929(5) & 6.44(2.71) & 31(10) & 4.94(27) & - & 769(88) & - & - & - & 24.52(8) & 769(88) & 902 & 2.94(28)\\\\
J1023$-$5746 & 55043(8) & - & - & - & - & - & 15.4(4) & 187(10) & 550(12) & 6.55(54) & 1110(31) & 961(16) & 2.19(7)\\\\
                        & 56004(8) & 65(19) & 112(32) & - & - & - & 9.97(72) & 66(4) & 422(5) & 12.6(7) & 620(15) & 643(16) & 2.26(10)\\\\
                        & 56647(8) & 11.8(10.0) & 37(30) & - & - & - & 16.8(1.1) & 205(15) & 487(14) & 6.09(1.16) & 1102(47) & 1080(16) & 2.30(16)\\\\
                        & 57727(8) & 21.7 & 13 & 5.51 & 59 & 484 & - & - & - & 23.4 & 484 & 475(25) & 2.89\\\\
                        & 58202(17) & - & - & 7.23(1.88) & 91(18) & 117(22) & 2.34(19) & 29(1) & 462(2) & 12.4(7) & 548(5) & 621(25) & 2.2(2)\\\\
                        & 58823(8) & - & - & 5.30(45) & - & 862(97) & 1.45(63) & 25(32) & 111(27) &0.99(67) &862(97)& - &0.77(10)\\\\
J2111$+$4606 & 55700(5) & - & - & 3.0(5) & - & 1864(90) & 2.28(1.83) & 127(111) & 248(127) & 5.03(51) & 1864(90) & 1862(20) & 1.03(20)\\\\
                         & 57562(15) & - & - & 3.53(47) & - & 2140(311) & 4.19(2.53) & 143(55) & 187(72) & 16.0(2.2) & 2140(311) & - & 2.37(34)\\\\
J2229$+$6114 & 55134(1) & - & - & 3.74(41) & - & 452(28) & 3.62(79) & 24(9) & 94(7) & 0.58(38) & 452(28) & 460(12) & 0.79(9)\\\\
                         & 55601(2) & 29.5(14.6) & 84(36) & - & - & - & 7.96(1.50) & 86(19) & 564(23) & 8.63(1.58) & 821(61) & 760(10) & 1.66(22)\\\\
                         & 56797(5) & - & - & - & - & - & 10.5(3) & 115(5) & 452(7) & 3.89(32) & 795(17) & 975(8) & 1.44(4)\\\\
                         & 58425(2) & 22.7(14.2) & 89(53) & - & - & - & 11.0(4) & 115(8) & 512(11) & 5.2(4) & 857(27) & 845(4) & 1.62(6)\\\\
\hline\\
\label{modelfit}
\end{tabular}}
\end{center}
\end{table*}

\section{Results on Neutron Star Structure} \label{sec:results} 

In order to compare the theoretical non-linear creep time-scale with the fit results on $\tau_{\rm nl}$ and $\tau_{\rm s}$, we use a neutron star model with radius $R=12$ km and mass $M=1.6M_{\odot}$.  For interior temperature, we employ the following formula which is relevant for cooling via modified URCA reactions \citep{yakovlev11}
\begin{equation}
T_{\rm in}= 1.78\times10^{8}~{\rm K}~\left(\frac{10^{4}{\rm yr}}{t_{\rm sd}}\right)^{1/6}.
\label{murca}
\end{equation}
\citet{erbil20} considered five crustal superfluid layers and estimated the theoretical range $\omega_{\rm cr}/E_{\rm p}\simeq0.5-0.01$\mevs,~with the lowest value is reached in the densest pinning region. Then, Eqs.(\ref{taunl}) and (\ref{murca}) yield
\begin{equation}
\tau_{\rm nl}\cong(30-1420)\left(\frac{\dot\nu}{10^{-11}\mbox{\hzs}}\right)^{-1}\left(\frac{t_{\rm sd}}{10^{4}{\rm yr}}\right)^{-1/6}\mbox{days},
\label{creepnl}
\end{equation}
which is monotonically decreasing towards the crust-core base. For the Vela pulsar, Eq.(\ref{creepnl}) estimates that the superfluid recoupling time-scale is $\tau_{\rm nl}=(19-807)$ days. This is consistent with the range $[\tau_{\rm nl}, \tau_{\rm s}]=(19-45)$ days from the fit results given in Table \ref{modelfit}. Since model fit results are towards the lower end of the theoretical estimate and encompass only a narrow interval, we conclude that the glitches of the Vela pulsar should start from the densest pinning regions of the crust and that entire crust is contributing to its glitch activity. This is consistent with the fact that the size distribution of Vela glitches does not change significantly even though waiting times between its glitches show deviations \citep{melatos18,fuentes19}. For PSR J1023$-$5746, Eq.(\ref{creepnl}) predicts $\tau_{\rm nl}=(9-375)$ days. This is consistent with the range $[\tau_{\rm s}]=(29-205)$ days from the fit results given in Table \ref{modelfit}. For PSR J2111$+$4606 $\tau_{\rm nl}=(48-2054)$ days interval is obtained which agrees with the values $\tau_{\rm s}=$(127 and 143) days from the fit results given in Table \ref{modelfit}. For PSR J2229$+$6114, Eq.(\ref{creepnl}) predicts the theoretically allowed interval $\tau_{\rm nl}=(10-434)$ days. This is consistent with the range $[\tau_{\rm nl}, \tau_{\rm s}]=(24-115)$ days from the fit results given in Table \ref{modelfit}.

\begin{figure}
\centering
\vspace{0.1cm}
\includegraphics[width=1.0\linewidth]{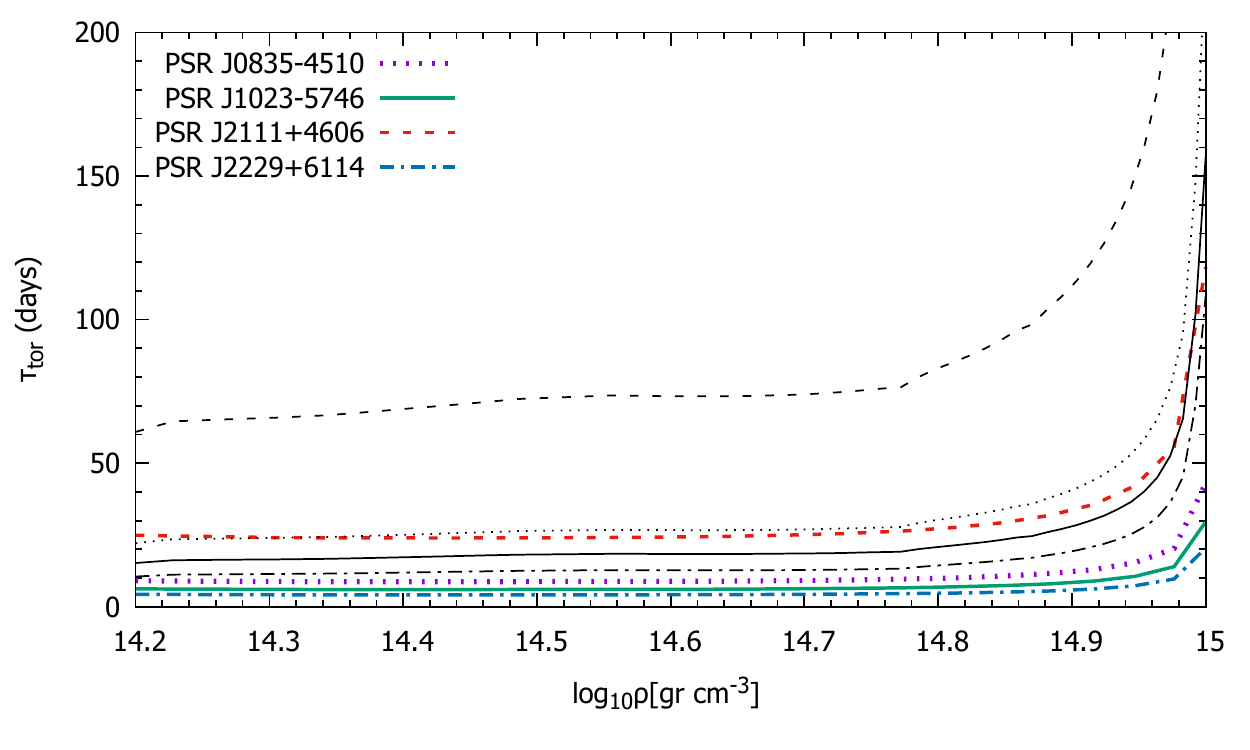}
\caption{Exponential decay time-scale of the toroidal field region in the neutron star core, Eq.(\ref{tautor}), versus matter density for Vela (dotted curves), PSR J1023$-$5746 (solid curves), PSR J2111$+$4606 (dashed curves) and PSR J2229$+$6114 (dotted-dashed curves). The black [coloured] curves correspond to \citet{akmal98}[\citet{douchin01}] EOS.}
\label{taucore}
\end{figure}

In Fig. \ref{taucore}, we show the relaxation time-scale of the toroidal field region, Eq.(\ref{tautor}), of four gamma-ray pulsars considered in this work versus core density for A18 + $\delta v$ + UIX* \citep{akmal98} and SLy4 \citep{douchin01} EOSs with microphysical parameters are taken from \citet{chamel08} and the adopted temperature evolution is given by Eq.(\ref{murca}). For the Vela pulsar exponential decay time constants are in the range $\tau_{\rm tor}=9(1)-31(10)$~days from fits [c.f. Table \ref{modelfit}] and in agreement with the neutron star's core response for \citet{douchin01} EOS. For PSR J1023$-$5746 fit results in Table \ref{modelfit} yield exponential decay times in the range $\tau_{\rm tor}=13-112(32)$~days which agrees well with the neutron star's core response for \citet{akmal98} EOS. For PSR J2229$+$6114  exponential decay times are $\tau_{\rm tor}=84(36)$ and $89(53)$~days from fit results in Table \ref{modelfit} which can be explained by \citet{akmal98} EOS.

\citet{alpar94} analysed the pulsar data and concluded that $\delta\Omega_{\rm s}/\Omega$ should be a constant within a factor of a few, reflecting the fact that crustal structure and deposited angular momentum are almost the same for pulsars. This prescription along with Eq.(\ref{offsettime}) leads to the following simple estimate for the interglitch time in terms of the parameters of the Vela pulsar \citep{alpar06}
\begin{equation}
t_{\rm g}=\frac{\delta\Omega_{\rm s}}{\Omega}\frac{\Omega}{\abs{\dot\Omega}}\cong2\left<\frac{\delta\Omega_{\rm s}}{\Omega}\right>t_{\rm sd}\cong3.5\times10^{-4}t_{\rm sd}.
\label{gtime}
\end{equation}
Eq.(\ref{gtime}) gives the estimate $t_{\rm g}=588$ days for typical time between the glitches of PSR J1023$-$5746 which is in agreement with the mean inter-glitch time $t_{\rm ig}=629\pm173$ days observed for this pulsar. For PSR J2111$+$4606 the estimate $t_{\rm g}=2237$ days is close to the observed time 1860(20) days between G1 and G2, and in agreement with the prediction of $t_{\rm ig}=2140(311)$ days for the next glitch from G2 within errors. For PSR J2229$+$6114 $t_{\rm g}=1342$ days is obtained which compares well with the mean inter-glitch time $t_{\rm ig}=1067\pm407$ days observed for this pulsar within errors.

\begin{figure}
\centering
\vspace{0.1cm}
\includegraphics[width=1.0\linewidth]{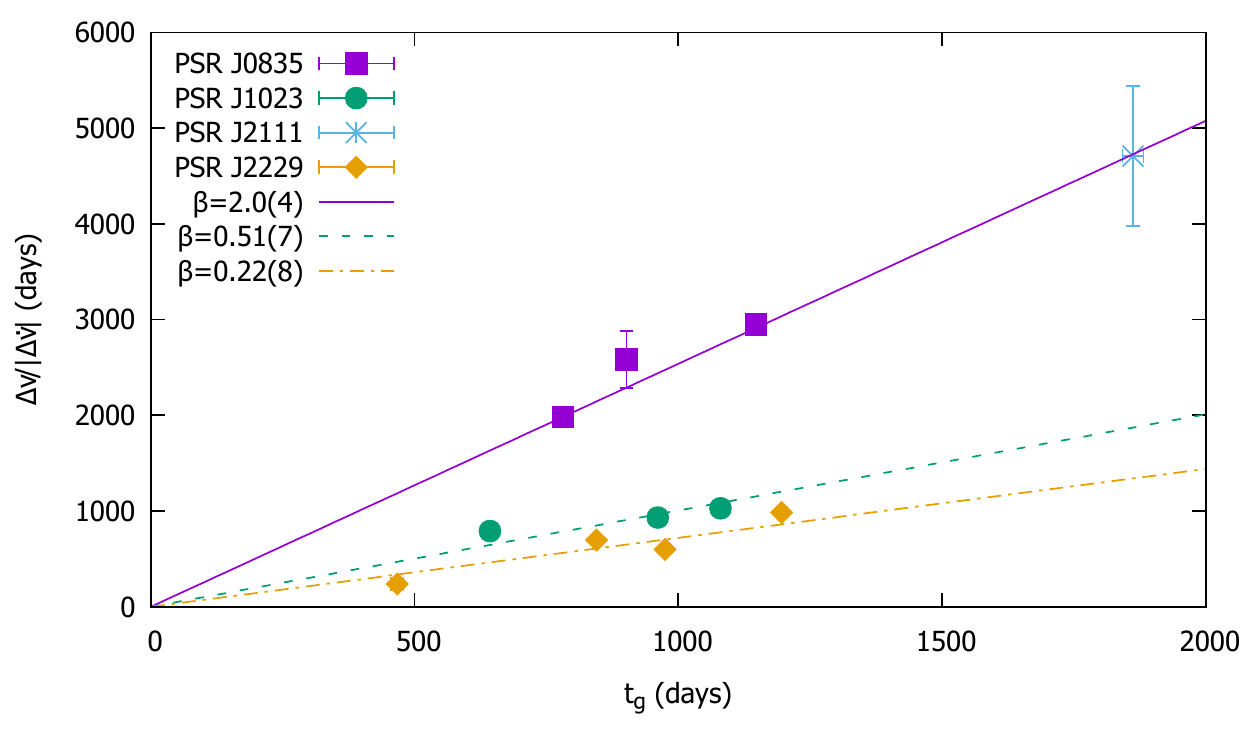}
\caption{Correlation corresponding to Eq.(\ref{creepcor}) for the glitches of four gamma ray pulsars analysed in this paper.}
\label{correlationglitch}
\end{figure}

According to the vortex creep model from equations (\ref{glitchdotnu}) and (\ref{glitchomega}) a correlation exists between the ratio of glitch magnitudes in frequency and spin-down rate $\Delta\nu/|\Delta\dot\nu|$ and inter-glitch time $t_{\rm g}$
\begin{equation}
\frac{\Delta\nu}{|\Delta\dot\nu|}=\left(\frac{1}{2}+\beta\right)t_{\rm g}.
\label{creepcor}
\end{equation}
The correlation coefficient $\beta=I_{\rm B}/I_{\rm A}$ gives a measure of the extend of the vortex traps within the neutron star crust. For the glitches of four gamma-ray pulsars analysed in this paper, the data corresponding to Eq.(\ref{creepcor}) are shown in Figure~\ref{correlationglitch}. The Vela and  PSR J2111$+$4606 lie on the same line, implying that vortex traps leading to their glitches have nearly the same structure and distribution. For PSR J1023$-$5746 and PSR J2229$+$6114, $\beta$ values are lower which means that these pulsars are in the process of developing vortex traps and will exhibit larger glitches in the future. For comparison, PSR J0537--6910 has $\beta=13.9(6)$ \citep{akbal21}.

In the vortex creep model, superfluid contribution to the glitch recovery process ends when the change in the superfluid angular velocity $\delta\Omega_{\rm s}$ due to vortex discharge is renewal by the slow down torque, i.e. after the offset time $t_{\rm g}=\delta\Omega_{\rm s}/\abs{\dot\Omega}$ elapses. $I_{\rm A}$, moment of inertia of the non-linear crustal superfluid regions, encodes the number of the vortex lines which gives rise to the glitch and in turn determines $\delta\Omega_{\rm s}$. Therefore, a correlation between $I_{\rm A}$ and $t_{\rm g}$ is expected on theoretical grounds \citep{alpar81,erbil20}: 
\begin{equation}
t_{\rm g}\sim f_{\rm v}\left(\alpha,\eta\right)t_{\rm sd}\left(\frac{D}{R}\right)\sim f_{\rm v}\left(\alpha,\eta,\theta_{\rm cr}\right)t_{\rm sd}\frac{I_{\rm A}}{I},
\label{tgmoi}
\end{equation}
where $t_{\rm sd}$ is the spin-down age, $R$ is the neutron star radius, $D$ is the broken crustal platelet size, $f_{\rm v}$ is a factor which depends on the inclination angle between the rotation and magnetic moment axes $\alpha$, the critical strain angle for crust breaking $\theta_{\rm cr}$, and vortex scattering coefficient $\eta$ that determines the tangential motion of the vortex lines between adjacent vortex traps. Therefore, $f_{\rm v}$ can be regarded as a structural parameter specific to a given pulsar. Theoretical estimation of $\theta_{\rm cr}\sim 10^{-2}-10^{-1}$ was obtained for neutron star matter \citep{baiko18}. The plate size is related to the critical strain angle via $D\sim\theta_{\rm cr}\ell_{\rm crust}$ with $\ell_{\rm crust}\approx0.1R$ being crust thickness \citep{akbal15,erbil19}. An order of magnitude estimate for $f_{\rm v}\left(\alpha,\eta,\theta_{\rm cr}\right)$ is $\sim10^{-2}-10^{-1}$.    
For the 2010 and 2013 Vela glitches $I_{\rm A}/I\cong6\times10^{-3}$ and $t_{\rm obs}\cong1200$ days which is consistent with the correlation given by Eq.(\ref{tgmoi}). On the other hand, for the 2016 and 2019 Vela glitches $I_{\rm A}/I\sim4\times10^{-3}$ and $t_{0}\sim800$ days which is also in line with our expectation. Even though only single crustal superfluid region takes part in its glitches, the relation $I_{\rm s}\propto t_{\rm s}$ is also satisfied for PSR J1023$-$5746 confirming the characteristic feature that waiting time between glitches not only scales inversely with $\abs{\dot\nu}$, but also depends linearly on $\delta\Omega_{\rm s}$.

Due to the crustal entrainment effect moments of inertia of the superfluid regions inferred from the torque equilibrium on the neutron star in the post-glitch relaxation phase should be multiplied by the enhancement factor $\langle m_{\rm n}^{*}/m_{\rm n}\rangle=5$ \citep{chamel17}. So, when reassessing a glitch one has to take into account of modification $\left[\langle m_{\rm n}^{*}/m_{\rm n}\rangle\left(I_{\rm A}/I\right)\right]$ for non-linear creep regions which are responsible for the vortex unpinning avalanche. Then, for the four Vela glitches considered here, the entrainment effect corrected total crustal superfluid moment of inertia participated in glitches is $(I_{\rm cs}/I)_{\rm entrainment}\leq4.5\times10^{-2}$ and can be accommodated with even for a $1.4M_{\odot}$ neutron star model \citep{delsate16}. For PSR J1023$-$5746 $(I_{\rm cs}/I)_{\rm entrainment}\leq7.5\times10^{-2}$ and can be accommodated with a neutron star model having a thick crust and stiff equation of state \citep{basu18}. This implies that PSR J1023$-$5746 should have a larger angular momentum reservoir and a slightly different internal structure from that of the Vela pulsar. While for PSR J2111$+$4606 and PSR J2229$+$6114 $(I_{\rm cs}/I)_{\rm entrainment}$ is $\leq5\times10^{-2}$ and $\leq6\times10^{-2}$, respectively and can be accommodated with a canonical neutron star model very close to that of the Vela pulsar.

\section{Discussion and Conclusions} \label{sec:dandc}  

In this paper, we have analysed the timing data from the Fermi-LAT observations of gamma-ray pulsars PSR J0835$-$4510 (Vela), PSR J1023$-$5746, PSR J2111$+$4606, and PSR J2229$+$6114. We identified 20 glitches, of which 11 are new discoveries (see Table \ref{glitchobs}). Our sample includes 15 large glitches with magnitudes $\Delta\nu/\nu\gtrsim10^{-6}$. We have done post-glitch timing fits within the framework of the vortex creep model for 16 glitches with $\Delta\nu/\nu\gtrsim10^{-7}$. Fit results are shown in Figs. \ref{vela2010}$-$\ref{J2229G8} and the inferred model parameters are given in Table \ref{modelfit}. For each case, we obtained moments of inertia of the various superfluid layers that contributed to the angular momentum exchange as well as total crustal superfluid moment of inertia involved in these glitches. We have also made a comparison between theoretical estimation for the time to the next glitch from the parameters of the previous glitch (which are deduced from the fit results) and the observed inter-glitch time-scales. It is found that the vortex creep model predictions are in qualitative agreement with the observed times. Note that one can make a rough estimate for the time of the next glitch of a pulsar from just two observables of its previous glitch as $t_{\rm g}\approx\Delta\dot\nu_{\rm p}/\ddot\nu_{\rm p}$, namely  in terms of the long-term step increase in the spin-down rate $\Delta\dot\nu_{\rm p}$ and the measured inter-glitch frequency second derivative $\ddot\nu_{\rm p}$ after all exponential recoveries are over without doing a detailed fit to the data. 

Our analyses revealed the detailed 4 post-glitch spin evolution of the well-studied Vela pulsar. We present the data on the post-glitch recovery phase of the 2019 Vela glitch for the first time. 

PSR J1023$-$5746 underwent 6 large glitches in a period of 12.5 years analysed in this work with glitch activity parameter $A_{\rm g}=14.5\times10^{-7}$\yr,~making this pulsar a promising candidate for frequently glitching Vela-like pulsars and an important target for future observations. Its glitch magnitudes are somewhat larger than that of the Vela pulsar. With an age of 4.6 kyr, PSR J1023$-$5746 is the most frequently glitching young pulsar displaying large glitches with $\Delta\nu/\nu\gtrsim3\times10^{-6}$. This pulsar is in the fourth rank among young neutron stars population in terms of its glitch magnitudes after the objects 0.8 kyr old PSR J1846$-$0258 \citep{livingstone10}, 1.6 kyr old PSR J1119$-$6127 \citep{weltevrede11,archibald16} and 7.4 kyr old PSR J1614$-$5048 \citep{yu13}, each of which exhibited sizeable glitches of $\Delta\nu/\nu\simeq6\times10^{-6}$. We anticipate that PSR J1023$-$5746 will continue to have large glitches and that both its glitch activity and glitch sizes tend to increase with time. The post-glitch recovery of PSR J1023$-$5746 is consistent with response of a single superfluid region of high enough moment of inertia [Eq.(\ref{creepsingle})] in contrast to the behaviour of the Vela pulsar in which integrated response [Eq.(\ref{creepfull}) or Eq.(\ref{glitchdotnu})] is seen.

We have identified the first two glitches of PSR J2111$+$4606. These glitches have common property that in their post-glitch recovery phase the responses of both integrated superfluid regions and single superfluid layer are superimposed. It seems that for the glitches of PSR J2111$+$4606 vortex lines in a single superfluid layer with high enough moment of inertia start an avalanche. This avalanche partakes vortices in the other crustal superfluid regions. Its response still can be seen appreciably because the radial extent of the single superfluid layer is large to be recognized within the data. Given that its glitch magnitudes and age are similar to that of the Vela pulsar we expect PSR J2111$+$4606 to experience numerous Vela like glitches in the future with somewhat longer repetition time.

By exhibiting 9 glitches in the range $\Delta\nu/\nu\approx6\times10^{-10}-1.5\times10^{-6}$ PSR J2229$+$6114 resembles the glitch activity of the PSR B1737$-$30 which experienced 36 glitches of magnitudes in the range $\Delta\nu/\nu\approx7\times10^{-10}-3\times10^{-6}$ \citep{liu19}. PSR J2229$+$6114 has an age very similar to that of the Vela pulsar while its spin-down rate is about twice that of the Vela. Then, our expectancy is that the typical time between the Vela and PSR J2229$+$6114 glitches scales inversely with the ratio of their spin-down rates, which is indeed the case. Therefore, we conclude that PSR J2229$+$6114 is a representative of Vela-like pulsars in many respects. The only distinction from the Vela pulsar's behaviour being the form of post-glitch recovery trend which is better described by the response of a single superfluid region like the case of PSR J1023$-$5746.

Observations of the post-glitch recovery of the spin-down rate allow us to pinpoint the location of the glitch trigger and yield the time-scale of the coupling of the superfluid regions participated in glitch to the other parts of the neutron star. The superfluid recoupling time-scale $\tau_{\rm nl}$ decreases with increasing density in the crust \citep{erbil20} and in large glitches for which glitch trigger is located deeper in the crust, $\tau_{\rm nl}$ is expected to be smaller. And this is indeed the observed case (see Eq.(\ref{creepnl}), Table \ref{modelfit}, and discussion in Section \ref{sec:results}). 

We determine braking index $n=$2.63(30) for PSR J2229$+$6114 after glitch contributions have been removed. A plasma-filled magnetosphere under central dipolar magnetic field approximation leads to braking index in the range $3\leq n\leq 3.25$ depending on the inclination angle \citep{arzamasskiy15,eksi16}. Then, there must be some mechanisms interior to the neutron star or operating in the magnetosphere to effectively reduce $n$ to 2.63 for PSR J2229$+$6114. Many different mechanisms can be responsible for the braking indices less than 3 \citep{blandford88}. Magnetic field related models refer to increase in the dipolar component as a result of rediffusion of the field lines which submerged by accretion of the fallback material onto the neutron star after supernova explosion \citep{geppert99} or conversion of the interior toroidal field component to poloidal one through the Hall drift effect \citep{gourgouliatos15}. External models invoke decelerating torque arising from pulsar winds \citep{tong17} or putative fallback disk \citep{menou01,benli17} or inclination angle growth due to the flow of magnetospheric charges in excess of the Goldreich-Julian value \citep{beskin84}. Another possibility is finite-size inertia effects of the near zone corotating plasma in the magnetosphere \citep{melatos97}. The measured braking index value for the gamma-ray pulsar PSR B0540$-$69 displayed significant variations after one of its glitches \citep{wang20}. Gamma-ray pulsar PSR J1124$-$5916 exhibited two state spin-down behaviour which underwent transition very close to its glitch with a low braking index $n=1.98$ \citep{ge20b}. Future timing and spectral observations of PSR J2229$+$6114 will help to constrain whether this pulsar displays braking index alternation in the aftermath of glitches like PSR B0540$-$69 and PSR J1124$-$5916 had suffered and shed light on the physical mechanisms responsible for deviation from $n\sim3$.

Several glitches have been observed from gamma-ray pulsars with the Fermi-LAT but they did not concurrent with detectable variability in $\gamma$-ray emission so far \citep{acero15}. If observed at all, any variability in $\gamma$-rays coincident with glitches has the potential to probe into large-scale changes in the current circulation and pair production level in the magnetosphere. Even though 1975 and 1981 glitches of the Vela pulsar were observed only in radio wavelengths, quite interestingly, there was a clear wandering of one of the pulse components detected in gamma rays \citep{grenier88}. This fact is an indicative of a global magnetospheric reconfiguration coincident roughly with the glitch dates. $\gamma$-ray variability was in the form of a jump with rise time $11.4\pm 1.0$ days followed by a gradual decay on a time-scale of $137\pm 39$ days, similar to the overtones of the observed post-glitch exponential decay time-scales due to the superfluid internal torques. If some part of the energy released inside the crust after a quake is pumped into the magnetosphere, several type short term mode changes can be induced in the magnetosphere \citep{yuan20}. Future observations of the glitching gamma-ray pulsars will help us to understand whether there exists an intimate relation between post-glitch exponential decay time-scales and $\gamma$-ray variability duration.  

Gamma rays are produced in the magnetospheres of pulsars as a result of the dynamics of magnetic field lines close to the light cylinder. Within their magnetospheres processes like partial twist or possibly small-scale reconnection of the magnetic field lines may occur. We speculate that gamma-loud pulsars would be more active glitchers due to the conveyed stress of magnetic field lines to the conducting crust via their anchored footpoints which will eventually strain the crust beyond the yield point along with the contribution from the spin-down and easily initiate large scale vortex unpinning avalanche. 

\section*{Acknowledgements}
\addcontentsline{toc}{section}{Acknowledgements}
The work of EG is supported by the Scientific and Technological Research Council of Turkey
(T\"{U}B\.{I}TAK) under the grant 117F330. The authors thank the National Natural Science Foundation of China for support under grants No. U1838201, U1838202, U1838104, U1938103, U1938109  and 11873080. We thank referee for useful comments which led to improvement of presentation.

\section*{Data Availability}

The data underlying this article will be shared on reasonable request to the corresponding authors.




\appendix

\section{Comparison with Other Glitching Gamma-Ray Pulsars} \label{sec:comparison}

In this appendix, we quote the glitch magnitudes and some distinctive features of the other gamma-ray pulsars from the published literature, which are summarized in Table \ref{tab:litglitch}. 

Gamma-ray pulsars obey the general trend of pulsar glitches mentioned in Section \ref{sec:creepmodel}: A bimodal distribution with a narrow peak toward larger glitch sizes, glitch activity increases with spin-down rate $A_{\rm g}\propto\abs{\dot\nu}$, and glitches are observed from all ages with large ones predominantly seen from young and mature objects. It seems that radio-loud gamma ray pulsars undergo systematically larger glitches compared to the radio-quiet ones. This can be due to the selection effects. Future observations will discriminate whether this is a general trend intrinsic to pulsar structure. 

There are some qualitative differences between radio-quiet and radio-loud isolated gamma-ray pulsars. In the radio-loud objects, the magnetic field at the light cylinder $B_{\rm LC}$ exhibits a Gaussian distribution, while in the radio-quiet objects  $B_{\rm LC}$ has a nearly uniform distribution \citep{malov15}. The radio-loud group has higher $B_{\rm LC}$ which can be understood in terms of their periods, since $B_{\rm LC}\sim B_{\rm s}P^{-3}$. Rotation period $P$ of radio-loud pulsars is generally smaller than the period of radio-quiet pulsars while the distribution of their surface magnetic field $B_{\rm s}$ does not differ significantly. Also, a certain correlation between their $\gamma$-ray luminosity  $L_{\gamma}$ and $B_{\rm LC}$ is found, which suggests that the radiation is generated in the vicinity of the light cylinder, probably due to the synchrotron mechanism \citep{malov15}. The observed much wider beam shapes in $\gamma$-rays as well as simulations support that $\gamma$-rays are produced in the outer magnetospheric gaps \citep{takata11,watters11}. Moreover, radio-loud gamma-ray pulsars dominate at higher $\dot E_{\rm sd}$ and cutoff energy $E_{\rm cutoff}$ strongly correlates with $B_{\rm LC}$ in radio quiet objects \citep{hui17}. By comparing the radio profile widths of  gamma ray pulsars and non gamma ray-detected pulsars, \citet{rookyard17} concluded that radio-loud gamma-ray pulsars typically have narrower radio beams. About 25\% of the known gamma-ray pulsars \footnote{https://confluence.slac.stanford.edu/display/GLAMCOG/Public+List+of
+LAT-Detected+Gamma-Ray+Pulsars.} are radio quiet while \citet{sokolova16} have estimated that the actual fraction of the radio-quiet gamma-ray pulsars may be as large as 70\% after selection effects are taken into account, which is again in favour of the outer gap magnetosphere models. It should also be noted that $\gamma$-ray-to-X-ray flux ratios of the radio-quiet pulsars are higher compared to radio-loud objects \citep{marelli15}.

\begin{table*}
\caption{Glitches and physical properties of other gamma-ray pulsars published in the literature.}
\label{tab:litglitch}
\begin{center}
\begin{threeparttable}
{\scriptsize
\begin{tabular}{lccccccccccD{.}{.}{1}D{.}{.}{0}D{.}{.}{4}l}
\hline\\
\multicolumn{1}{c}{Pulsar} & \multicolumn{1}{c}{Age} & \multicolumn{1}{c}{$|\dot\nu|$} & \multicolumn{1}{c}{$\dot E_{\rm sd}$} & \multicolumn{1}{c}{$B_{\rm s}$} &  \multicolumn{1}{c}{$B_{\rm LC}$} &  \multicolumn{1}{c}{Radio State} & \multicolumn{1}{c}{Glitch Date} & $\Delta\nu/\nu$ & $\Delta\dot\nu/\dot\nu$ & \multicolumn{1}{c}{Reference} \\
& \multicolumn{1}{c}{($10^{3}$\,yr)} & \multicolumn{1}{c}{($10^{-12}$\hzs)} & \multicolumn{1}{c}{($10^{35}$\ergs)} & \multicolumn{1}{c}{($10^{12}$\,G)} & ($10^{3}$\,G) &  & (MJD) &  ($10^{-9}$) & ($10^{-3}$) & \\ 

\hline\\
J0007$+$7303 & 13.9 & 3.6 & 4.5 & 11 & 3.21 & Quiet & 54953 & 553(1)  & 1.0(2) & \citet{ray11} \\
 & & & & & & & 55466 & 1260 & -& \citet{espinoza11} \\\\
J0205$+$6449 & 5.37 & 45 & 270 & 3.61 & 119 & Loud & 52555(8) & 340(110) & 5(1) & \citet{espinoza11} \\
 & & & & & & & 52920(72) & 3800(400) &12(1)& \citet{espinoza11} \\
 & & & & & & & 54907(3) & 1733(6) &6(1)& \citet{espinoza11} \\
 & & & & & & & 55275(1) & 3.8(2) &-0.92(2)& \citet{espinoza11} \\
 & & & & & & & 55350.3(7) & 6.9(3) &-0.54(4)& \citet{espinoza11} \\
 & & & & & & & 55416(1) & 4.9(4) &-0.76(4)& \citet{espinoza11} \\
 & & & & & & & 55511(2) & 10.6(8) &0.6(2)& \citet{espinoza11} \\
 & & & & & & & 55721.56(13) & 118.8(10.7) &2.65(6)& \citet{espinoza11} \\
 & & & & & & & 55816(8) & 3005(40) &18.8(3)& \citet{espinoza11} \\
 & & & & & & & 56835(24) & 87(11) &2.6(2)& \citet{espinoza11} \\
 & & & & & & &57358(1) & 531(4) &6(1)& \citet{espinoza11} \\
 & & & & & & &57697(1) & 41.9(1.4) &-& \citet{espinoza11} \\
 & & & & & & &58321(99) & 3162(3) &20.4(1)& \citet{espinoza11} \\\\
J0633$+$1746 & 342 & 0.195 & 0.32 & 1.6 & 1.15 & Faint & 50382 & 0.62 & - & \citet{jackson02} \\\\
J1048$-$5832& 20.4&6.28 & 20 & 3.49 & 17.3 & Loud & 48946.9(2) & 17.95(19) & - & \citet{yu13} \\
 & & & & & & &49034(9) & 2995(26) &-& \citet{yu13} \\
 & & & & & & &50791.485(5) & 768(3) &3.7(8)& \citet{yu13} \\
 & & & & & & &52733(37) & 1838.4(5) &3.7(3)& \citet{yu13} \\
 & & & & & & &53673.0(8) & 28.5(4) &0.19(14)& \citet{yu13} \\
 & & & & & & &54495(10) & 3042.56(14) &5.6(1)& \citet{yu13} \\\\
J1124$-$5916 & 2.85 & 41 & 120 & 10.2 & 38.5 & Faint & 55191 & 16(1) & 4.72(3)& \citet{ray11} \\
& & & & & & &58632(10) & 25(21) &0.77(13)& \citet{ge20b} \\\\
J1413$-$6205 & 62.8 & 2.3 & 8.3 & 1.76 & 12.5 & Quiet & 54735 & 1730& - & \citet{sazparkinson10} \\\\
J1420$-$6048& 13&17.9 & 100 & 2.41 & 71.3 & Loud & 51600(77) & 1146.2(6) & 3.83(8) & \citet{yu13} \\
 & & & & & & &52754(16) & 2019(10) &6.6(8)& \citet{yu13} \\
 & & & & & & &53725(9) & 1270(3) &3.9(3)& \citet{yu13} \\
 & & & & & & &54653(20) &934.5(4) &4.84(6)& \citet{yu13} \\
 & & & & & & &55410(19) & 1346.00(18) &-& \citet{yu13} \\
 & & & & & & &56270 & 1965 &-& \citet{lin21} \\
 & & & & & & &57230 & 1216 &-& \citet{lin21} \\
 & & & & & & &58560 & 1497 &-& \citet{lin21} \\\\
 J1709$-$4429& 17.5&8.86 & 34 & 3.12 & 27.2 & Loud & 48779(33) & 2050.6(4) & 5.86(8) & \citet{yu13} \\
 & & & & & & &51488(37) & 1166.73(17) &6.22(3)& \citet{yu13} \\
 & & & & & & &52716(57) & 2872(7) &8.0(7)& \citet{yu13} \\
 & & & & & & &54710(22) & 2749.7(1) &4.95(1)& \citet{weltevrede10} \\
 & & & & & & &58176(6) & 52.4(1) &-& \citet{lower18} \\\\
J1813$-$1246 & 43.4 & 7.6 & 62 & 0.93 & 78.5 & Quiet & 55094.1227 & 1166(0.4) & 6.4(3)& \citet{ray11} \\\\
J1838$-$0537 & 4.89 & 22.2 & 60 & 8.39 & 25.4 & Quiet & 55100 & 5500 & -15& \citet{pletsch12} \\\\
J1844$-$0346 & 11.6 & 12 & 42 & 4.23 & 27.6 & Quiet & 56135(7) & 3504(14) & 11.4(9)& \citet{clark17} \\\\
J1906$+$0722 & 49.2 & 2.88 & 10 & 2.02 & 13.7 & Quiet & 55067(7) & 4538(1) & 15.5(4)& \citet{clark15} \\\\
J1907$+$0602 & 19.5 & 7.63 & 28 & 3.08 & 23.8 & Faint & 55422 & 4657 & 13.3(1)& \citet{xing14} \\\\
J1952$+$3252 & 107 & 3.74 & 37 & 0.486 & 73.8 &Loud & 51967(9) & 2.25(9) &-0.2(1)& \citet{janssen06} \\
& & & & & & &52385(11) & 0.72(9) &-0.04(8)& \citet{janssen06} \\
& & & & & & &52912(5) & 1.29(7) &0.30(9)& \citet{janssen06} \\
& & & & & & &53305(6) & 0.51(9) &0.11(7)& \citet{janssen06} \\
& & & & & & &54103.44(3) & 5.2(2) &0.0(4)& \citet{espinoza11} \\
& & & & & & &55328(1) & 1500(1) &28(1)& \citet{espinoza11} \\\\
J2021$+$3651& 17.2&8.9 & 34 & 3.19 & 26.8 & Loud & 52630.1(1) &2587(2) & 6.2(2) & \citet{espinoza11} \\
& & & & & & &54177(25) & 745(6) &5.5(1)& \citet{espinoza11} \\
& & & & & & &55110(1) & 2230(1) &9.9(4)& \citet{espinoza11} \\
& & & & & & &57204.89(2) &3048(1) &6.68(31)& \citet{espinoza11} \\
 & & & & & & &58668(1) & 1186.5(8) &4.63(6)& \citet{espinoza11} \\\\

\end{tabular}}
\end{threeparttable}
\end{center}
\end{table*}


\bsp	
\label{lastpage}
\end{document}